\documentclass[a4paper,11pt]{article}
\pdfoutput=1
\usepackage{amssymb,mathrsfs}
\usepackage{bbm,bm}
\usepackage{siunitx}
\usepackage{amsmath}
\usepackage{braket}
\usepackage{mathtools}
\usepackage{color}
\usepackage{lmodern}
\usepackage{footnote}
\usepackage{slashed}
\usepackage{mathrsfs}
\usepackage[pdftex] {graphicx}
\usepackage{multirow}
\usepackage{jheppub} 
\usepackage{here}
\usepackage[justification=centering, singlelinecheck=false]{caption}
\usepackage[list=true, labelfont=bf, labelformat=brace, position=top]{subcaption}

\usepackage[utf8]{inputenc} 
\usepackage{hyperref}
\usepackage{orcidlink}

\usepackage{xcolor}
\definecolor{greena}{rgb}{0.0, 0.5, 0.0}

\graphicspath{{newfig/}}
\renewcommand\arraystretch{1.25}
\allowdisplaybreaks

\newcommand{\eps}{{\varepsilon}}

 \author[a,e]{Yong-Hui Lin,}
 \author[a,b]{Hans-Werner~Hammer,}
 \author[c,d,e]{and Ulf-G.~Mei{\ss}ner}
 
\affiliation[a]{Institut für Kernphysik, Technische Universität Darmstadt, 
     64289 Darmstadt, Germany}
\affiliation[b]{ExtreMe Matter Institute EMMI and Helmholtz Forschungsakademie Hessen f\"ur FAIR (HFHF), GSI Helmholtzzentrum 
 für Schwerionenforschung GmbH, 64291 Darmstadt, Germany}
\affiliation[c]{Helmholtz--Institut f\"ur Strahlen- und Kernphysik (Theorie)\\ 
 and Bethe Center for Theoretical Physics, Universit\"at Bonn, D-53115 Bonn, Germany}
\affiliation[d]{Institute for Advanced
 Simulation (IAS-4), 
 Forschungszentrum J\"ulich, D-52425  J\"ulich, Germany}
\affiliation[e]{Peng Huanwu Collaborative Center for Research and Education, International Institute for Interdisciplinary and Frontiers, Beihang University, Beijing 100191, China}

\emailAdd{yonghuil@buaa.edu.cn}
\emailAdd{Hans-Werner.Hammer@physik.tu-darmstadt.de}
\emailAdd{meissner@hiskp.uni-bonn.de}

\title{Short-range production of three bottom mesons }

\abstract{
    Previous investigations of the three-body dynamics of $B$ mesons
    have shown that no Efimov effect arises in systems composed of three $B$ and $B^*$ mesons. 
    This implies that the properties of such three-body systems can be described reliably within nonrelativistic effective field theory (NREFT) with short-range interactions using only two-body input, as three-body forces are strongly suppressed. 
    In this work, we present leading-order predictions for the three-body point production rates of systems consisting of three $B$ and $B^*$ mesons. 
    These predictions provide a novel way to experimentally probe the $B^{(*)}$-$\bar{B}^{(*)}$ interactions, which play a crucial role in the hadronic-molecule interpretation of the $T_{b\bar{b}1}(10610)$ and $T_{b\bar{b}1}(10650)$ states.
    Moreover, they provide a way to test the approximate conformal symmetry predicted for such systems at low energies experimentally.
}

\begin{document}
\maketitle
\flushbottom

\section{Introduction}
Since 2003, when the first signal of a new hadron with exotic nature was established through the observation of $X(3872)$ (also known as $\chi_{c1}(3872)$~\cite{ParticleDataGroup:2024cfk}) in exclusive $B^{\pm}\!\to\! K^\pm \pi^+\pi^- J/\psi$ decays by the Belle Collaboration~\cite{Belle:2003nnu}, a steadily growing family of exotic hadron candidates, lying beyond the conventional quark model of quark-antiquark mesons and three-quark baryons, has been reported experimentally and extensively investigated theoretically (see Refs.~\cite{Hosaka:2016pey,Esposito:2016noz,Guo:2017jvc,Olsen:2017bmm,Karliner:2017qhf,Kalashnikova:2018vkv,Brambilla:2019esw,Meng:2022ozq,Liu:2024uxn,Chen:2024eaq} for recent reviews).
Unraveling the internal structure of these exotic hadrons has become a central topic in hadron physics.
However, it remains challenging to unambiguously identify 
the nature of these states,
owing to the limited precision and diversity of current experimental probes. 
For instance, regarding the $X(3872)$, several competing hypotheses have been proposed: a hadronic molecule, that is a bound state of color-singlet mesons held together by residual strong forces~\cite{Close:2003sg,Voloshin:2003nt,Wong:2003xk,Braaten:2003he,Swanson:2003tb,Tornqvist:2004qy}, a compact tetraquark formed by tightly bound diquark-antidiquark pairs~\cite{Maiani:2004vq,Braaten:2024tbm,Brambilla:2024thx}, 
a gluonic hybrid state~\cite{Close:2003mb,Li:2004sta}, 
or even a conventional charmonium assignment~\cite{Eichten:2002qv,Eichten:2004uh}. 
To date, no consensus has been reached on which of these interpretations is correct, though
the hadronic molecular picture can most easily explain the existing data.

In the hadronic molecule scenario of exotic states,
the two-body interactions responsible for binding the constituents are known to leave clear imprints on three-body dynamics~\cite{Braaten:2004rn,Bedaque:1998kg,Bedaque:1998km,Bedaque:1999ve,Liu:2024uxn}, 
thereby offering a novel means to quantify the underlying two-body forces through suitable three-body observables.

The Efimov effect, a striking phenomenon in three-body dynamics, describes the emergence of shallow three-particle bound states (trimers) in systems with resonant interactions characterized by a large two-body scattering length $a$~\cite{Efimov:1970zz}. 
Given that the $Z_{b}(10610)$ and $Z^\prime_{b}(10650)$ (also referred to as $T_{b\bar{b}1}(10610)$ and $T_{b\bar{b}1}(10650)^+$, respectively~\cite{ParticleDataGroup:2024cfk}, and denoted hereafter as $Z$ and $Z^\prime$) lie very close to the $B^*\bar{B}$ and $B^*\bar{B}^*$ thresholds, respectively, the possible emergence of the Efimov physics in systems of three $B$ mesons has been investigated in detail within a dimer--particle framework based on nonrelativistic effective field theory, assuming that $Z$ and $Z^\prime$ are shallow bound states of the corresponding open-bottom channels~\cite{Lin:2017dbo}. 
The argument for the existence or non-existence of the Efimov effect is most straightforward for systems of identical bosons. In this case it occurs when at least two out of three pairs have resonant attractive interactions. In systems where the relevant two-body subchannels have
nontrivial spin and isospin quantum numbers and are superpositions of different pair interactions, like the $Z$ and $Z^\prime$, the counting is more complicated. Here one has to determine an effective number of attractively interacting pairs and it is more convenient to analyze the cutoff dependence of observables. 
The physical reason for the absence of the Efimov effect can be deduced from the $Z^{(\prime)}$-$B^{(*)}$ scattering lengths calculated in  \cite{Lin:2017dbo}. In some channels, the scattering length is small and positive, which indicates a weak repulsion in the three-body system. The $(I,S)=(3/2,1(2))$ and $(1/2,0)$ channels of the $Z^\prime_b B^*$ system, however,  have a large negative scattering length indicating an attractive interaction just below the critical 
attraction for the formation of an Efimov state. For a more detailed discussion of this issue, we refer the reader to Ref.~\cite{Lin:2017dbo}.  
Moreover, the absence of Efimov states implies that no three-body force is required at leading order in NREFT with only short-range contact interactions considered, which in turn enables parameter-free predictions for other three-body observables within the same framework. 
Similar analyses have also been performed for systems of $D$ and $D^*$ mesons~\cite{Canham:2009zq,Fu:2025joa}.

Another relevant process is short-range production, which has gained renewed interest since the proposal of the “unparticle'' concept in Ref.~\cite{Georgi:2007ek}, describing systems generated by local operators with nontrivial scale invariance. 
More recently, several physical realizations of nonrelativistic unparticles in nuclear and hadronic reactions have been proposed \cite{Hammer:2021zxb,Braaten:2021iot,Braaten:2023acw}. These studies demonstrate that a characteristic power-law behavior, governed by a definite scaling dimension, signals the emergence of an unparticle.
Such behavior can be observed in the invariant mass distribution of selected final states in short-range production processes (see Refs.~\cite{Hammer:2021zxb,Braaten:2021iot,Braaten:2023acw} for details).
A particularly notable feature of systems exhibiting unparticle-like behavior is a large scattering length in the two-body subsystems. This insight provides a means to extract two-body interactions by analyzing the corresponding short-range production of three-body systems.

In this work, we extend previous studies of the three $B$ mesons system. Rather than focusing on the search for Efimov effects, we investigate the possible emergence of unparticle behavior in this system by analyzing the short-range production of three $B$ mesons~\footnote{Note that the Efimov effect and unparticle behavior cannot emerge simultaneously in a three-body system. As pointed out in Refs.~\cite{Bedaque:1998kg,Braaten:2004rn}, the appearance of Efimov states requires an oscillatory three-body force, which explicitly breaks the approximate conformal symmetry that could otherwise give rise to unparticle behavior.}. 
This paper is organized as follows. In Section~\ref{sec:NREFT}, we briefly review the nonrelativistic effective field theory used to describe the three $B$ meson system. In Section~\ref{sec:ppd}, we present the formalism for short-range production in both the $ZB$ two-body and $BBB$ three-body systems. The derivation of the characteristic scaling dimension governed by approximate conformal symmetry is given in Section~\ref{sec:cft}. Numerical results and their analysis are presented in Section~\ref{sec:results}. Finally, we summarize our conclusions in Section~\ref{sec:end}.

\section{Dimer--particle interactions in NREFT \label{sec:NREFT}}
This section provides a short introduction to the fundamental theoretical tools used to analyze the three $B$ meson system in the low-momentum region, including the leading-order Lagrangian in NREFT, the dimer propagator, and the partial-wave projection.
 
\subsection{Effective Lagrangian and dimer propagator}
The question of whether the $Z$ and $Z^\prime$ mesons are virtual states, bound states or resonances has not been answered definitely.
Since a molecular bound state remains a viable option, we treat both the $Z$ and $Z^\prime$ with $I^G(J^{PC}) = 1^+(1^{+-})$ as $S$-wave hadronic molecules, i.e. shallowly bound states of two bottom mesons ~\cite{Baru:2017gwo,Lin:2017dbo}. 
The typical momentum scale of such a two-bottom-meson composite system is characterized by their binding momentum $\gamma = \sqrt{2 \mu E_B}$, where $E_B = M_{\rm threshold}-M_{Z}$ denotes the binding energy and $\mu$ is the reduced mass of the constituents.
In general, when the binding momentum is much smaller than the pion mass, $\gamma \ll M_\pi$, the constituent particles can be regarded as nonrelativistic and effectively point-like particles that interact only through short-range contact interactions. 
As a consequence, a nonrelativistic effective field theory containing only contact interactions at leading order (LO) provides a natural framework to describe the dynamics of such few-body systems. In the nuclear physics literature, this short-range EFT is also known as pionless EFT. It has been successfully applied to few-nucleon systems such as the deuteron and the triton~\cite{vanKolck:1997ut,vanKolck:1998bw,Kaplan:1998tg,Kaplan:1998we}.

Taking the binding energy $E_B=5$~MeV for the $Z$ and $E_B^\prime=1$~MeV for the $Z^\prime$, as adopted in Refs.~\cite{Baru:2017gwo,Lin:2017dbo},
together with the bottom-meson masses $M_B=5279.41$~MeV and $M_{B^*}=5324.75$~MeV from the last Review of Particle Physics (RPP)~\cite{ParticleDataGroup:2024cfk}, one obtains $\gamma=162$~MeV and $\gamma^\prime=72$~MeV\footnote{In principle, these binding energies could be fixed using the reported masses of the $Z$ and $Z^\prime$ in the RPP~\cite{ParticleDataGroup:2024cfk}, which spread around zero with substantial uncertainties. Here we instead adopt the heavy-quark effective theory determinations of Ref.~\cite{Cleven:2011gp}, where $Z$ and $Z^\prime$ are identified as shallowly bound two-bottom-meson states, consistent with our picture. }.
It follows that $\gamma^\prime/M_\pi\approx 0.5$, indicating that an NREFT without explicit pions should be reasonably applicable to the $Z^\prime B^*$.
In contrast, for the $Z$ state one finds $\gamma/M_\pi\approx 1$, suggesting that such a pionless description becomes marginal and potentially problematic.
Nevertheless, this NREFT framework may still be employed as a phenomenological model to gain first insights into the properties of the $ZB$ system, as done in Ref.~\cite{Lin:2017dbo}. 
A more accurate description of the $ZB$ dynamics would require the inclusion of pions as explicit degrees of freedom, where pion exchange provides the dominant finite-range interaction. Such an approach has been developed for the charm sector in Refs.~\cite{Fleming:2007rp,Braaten:2010mg}, but lies beyond the scope of the present work.
Instead, to assess the reliability of the leading-order NREFT, we investigate finite-range corrections to the LO prescription. These corrections represent a straightforward higher-order effect that can be addressed without fundamentally altering the underlying theoretical framework~\cite{Phillips:1999hh,Hammer:2001gh,Bedaque:2002yg,Afnan:2003bs,Braaten:2004rn,Ebert:2021epn}. 

We first write down the nonrelativistic effective Lagrangian at leading order for the dynamics of three bottom mesons~\cite{Lin:2017dbo,Wilbring:2016bda}
\begin{align}\label{eq: Lagrangian}
    &\mathcal{L} ={} B_{\alpha}^{\dagger} \left(i \partial_t + \frac{\nabla^2}{2M_B} \right) B_{\alpha} \: + \: \bar{B}_{\alpha}^{\dagger}
    \left(i \partial_t + \frac{\nabla^2}{2M_B} \right) \bar{B}_{\alpha} \notag\\
    &\phantom{xx} + \: B^{* \dagger}_{i \alpha} \left(i \partial_t + \frac{\nabla^2}{2M_{B^*}} \right) B^*_{i \alpha} \: + \: \bar{B}^{* \dagger}_{i \alpha}
    \left(i \partial_t + \frac{\nabla^2}{2M_{B^*}} \right) \bar{B}^*_{i \alpha} + \: Z^{\dagger}_{i A} \Delta Z_{i A} \: + \: Z^{^\prime \dagger}_{i A} \Delta^\prime Z^\prime_{i A} \nonumber\\
    &\phantom{xx} - \: g \Big[ Z^{\dagger}_{i A} \: \bigg(\bar{B}^*_{j \alpha} \: \delta_{ij} (\tau_2 \tau_A)_{\alpha \beta} \: B_{\beta} \: + \: \bar{B}_{\alpha}
    \: \delta_{ij} (\tau_2 \tau_A)_{\alpha \beta} \: B^*_{j \beta} \bigg) \: + \: h.c. \Big] \nonumber\\
    &\phantom{xx} - \: g^\prime \left[ Z^{^\prime \dagger}_{i A} \: \bar{B}^*_{j \alpha} \: (U_i)_{jk} (\tau_2 \tau_A)_{\alpha \beta} \: B^*_{k \beta} \: + \: h.c. \right]  \:.
\end{align}
Here, the lowercase Greek indices ($\alpha , \beta , \gamma... \in \{1,2\}$) label the isospin-$1/2$
components of the bottom mesons, while the uppercase Latin indices ($A,B,C... \in \{1,2,3\}$) denote the isospin components of the isospin-$1$ dimer fields.
The lowercase Latin indices ($i,j,k... \in \{1,2,3\}$) refer to spin-$1$ degrees of freedom.
The matrices $\tau_A$ are Pauli matrices acting in isospin space, and $U_i$ are the generators of the rotation group in the isospin-$1$ representation.
The last two lines introduce the coupling constants $g$ and $g^\prime$, which describe the interactions between the dimer fields and their constituent bottom mesons.
The coefficients $\Delta$ and $\Delta^\prime$ appearing in the kinetic terms of the $Z$ and $Z^\prime$
fields are also constants.
At LO, $\Delta^{(\prime)}$ and $g^{(\prime)}$ are not independent
and only the combinations $g^{(\prime)2}/\Delta^{(\prime)}$ enter physical observables.

\begin{figure*}[htb]
    \begin{center}
        \includegraphics[width=0.8\textwidth]{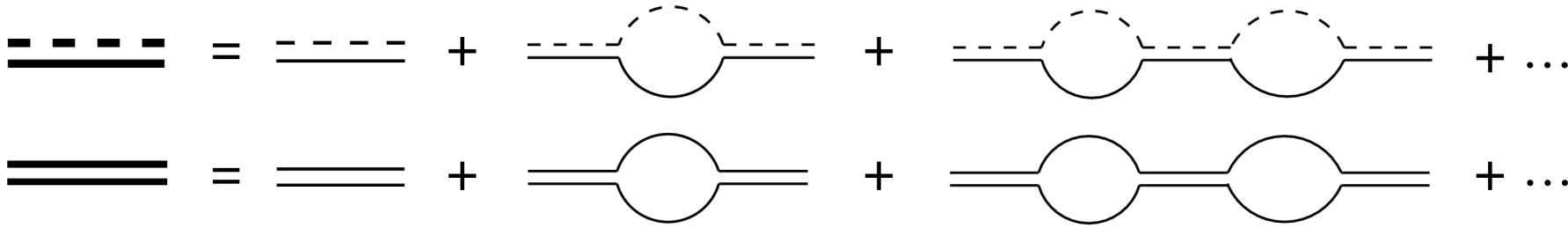}
    \end{center}
    \caption{\label{fig: self}{Diagrams illustrating the dressing of the dimer field. The dashed-solid and double-solid line are the $Z$ and $Z^\prime$ dimer propagators, respectively. The solid and dashed lines represent the $B^*$ and $B$ mesons, respectively. }}
\end{figure*}
In the LO Lagrangian of Eq.~\eqref{eq: Lagrangian}, the $Z$ field is treated as a dimer composed of $(B\bar{B}^*+\bar{B}B^*)/\sqrt{2}$, while the $Z^\prime$ field corresponds to a $B^*\bar{B}^*$ dimer. 
Both dimer fields are nondynamical, with their bare propagators dressed by bottom-meson loops, as illustrated in Figure~\ref{fig: self}.
To adopt a unified notation for different dimer states, we introduce a dimensionless parameter $r$ that characterizes the mass ratio of the constituent particles in a given dimer, defined as the ratio of the lighter constituent mass to the heavier one. Accordingly, one has $r=M_B/M_{B^*}=0.991485$ for the $Z$ dimer and $r^\prime=1$ for the $Z^\prime$ dimer. 
In the following, we set $M=M_B$, such that $M_{B^*}=M/r$. 
The corresponding reduced masses are
\begin{equation}
    \mu=\frac{M}{1+r},\quad \mu^\prime=\frac{M_{B^*}}{1+r^\prime}=\frac{M}{2r},
\end{equation}
respectively.
The propagators for the constituent particles and the dimers are then given by
\begin{align}\label{eq: propagators}
    &i S (p_0, \vec{p}) = \frac{i}{p_0-\frac{{p}^2}{2 M}+i\eps},\quad i S^* (p_0, \vec{p}) = \frac{i}{p_0-r\frac{{p}^2}{2 M}+i\eps},\notag\\
    &i D^{-1} (p_0, \vec{p}) = -i \frac{2 g^2 \mu}{\pi} \left[{-\frac1{a}+ \sqrt{\frac{r}{(1+r)^2}{p}^2 -2\mu p_0  -  i\eps}}\right]\:,\notag\\
    &i {D^\prime}^{-1} (p_0, \vec{p}) = -i \frac{2 g^{\prime 2} \mu^\prime}{\pi} \left[{-\frac1{a^\prime}+ \sqrt{\frac{r^\prime}{(1+r^\prime)^2}{p}^2 -2\mu^\prime p_0  -  i\eps} }\right]\:,
\end{align}
Here $p=|\vec{p}|$ denotes the magnitude of the three-dimensional space momentum $\vec{p}$, and $\eps \to 0^+$.
The unknown constants $\Delta^{(\prime)}$ in the dimer bare propagator are replaced by the inverse of corresponding two-body scattering length $a^{(\prime)}$, which are determined by matching the elastic tree-level amplitude to the leading-order effective range expansion (ERE)~\cite{Bedaque:1998kg,Lin:2017dbo,Wilbring:2016bda}. At this order, the two-body binding momentum $\gamma$ is directly related to the scattering length via $\gamma = 1/a$.
The wave-function renormalization constant $W$ for the dimer field
is then given by the residue at the bound-state pole of the corresponding propagator in Eq.~\eqref{eq: propagators}:
\begin{align}
    W = \frac{ \pi}{2 a  g^2 \mu^2 } \:,\quad W^\prime =\frac{ \pi}{2 a  g^{\prime 2} \mu^{\prime 2} } = \frac{2\pi r^2 }{ a^\prime g^{\prime 2} M^2} \:.
\end{align}

\subsection{Independent tree-level amplitudes of dimer--particle scattering}
\begin{figure*}[htb]
    \begin{center}
        \includegraphics[width=0.5\textwidth]{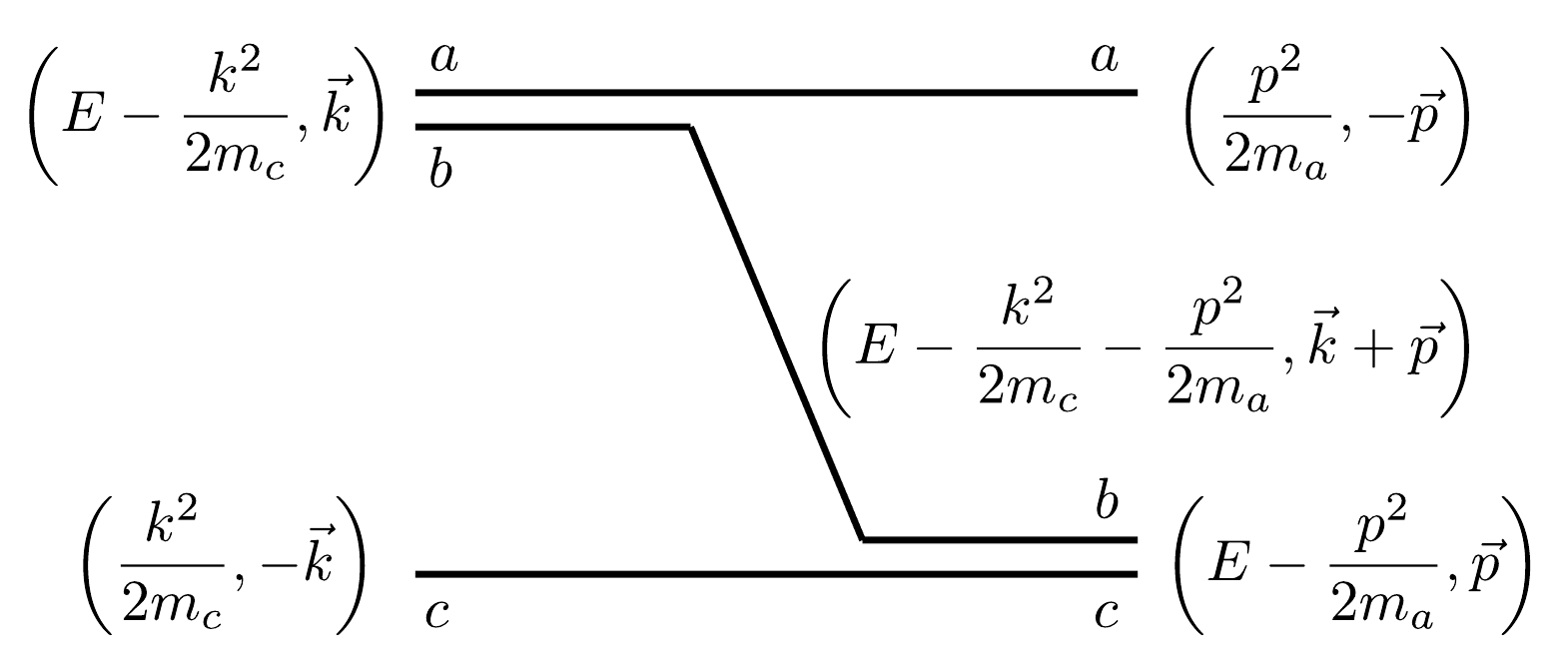}
    \end{center}
    \caption{\label{fig: kinematic}{Kinematic definitions for the $ZB$ scattering. 
            Energies and momenta assigned to the lines are given as (energy, momentum) with $E$ the center-of-mass energy. 
            The associated momenta can be used in
            all $ZB$ scattering processes. The labels $a$, $b$, and $c$ denote $B$ or $B^*$ mesons.
    }}
\end{figure*}
At leading order, the dimer--particle scattering amplitudes are given by the tree-level diagrams shown in Figure~\ref{fig: kinematic}. 
These amplitudes provide the basic kernels entering the integral equations used in the subsequent analysis.
For later convenience, we summarize in this subsection all independent tree-level amplitudes for dimer--particle scattering.
The relevant two-body channels are labeled as $1$--$4$, corresponding to $ZB$, $ZB^*$, $Z^\prime B$, and $Z^\prime B^*$, respectively.
From the LO Lagrangian in Eq.~\eqref{eq: Lagrangian}, dimer--particle scattering involves five independent scattering blocks, which are given by
\begin{align}\label{eq: tree_amp}
	&i{\cal M}_{11,\ i A \alpha }^{\phantom{xx}\ j B \beta}(E, \vec{k}, \vec{p})={} \frac{-i  g^2\delta_{ij}(\tau_A\tau_B)_{\beta\alpha}}{E-\frac{k^2}{2M}-\frac{p^2}{2M}-\frac{(\vec{k}+\vec{p})^2}{2M/r}+i\eps}\equiv \frac{-i 2 M g^2{\cal O}_{11,\ i A \alpha }^{\phantom{xx}\ j B \beta}}{2 M E-{k^2}-{p^2}-r(\vec{k}+\vec{p})^2+i\eps},\notag\\
	&i{\cal M}_{22,\ i m A \alpha }^{\phantom{xx}\ j n B \beta} (E, \vec{k}, \vec{p}) ={} \frac{-i  g^2\delta_{j m}\delta_{i n}(\tau_A\tau_B)_{\beta\alpha}}{E-\frac{k^2}{2M/r}-\frac{p^2}{2M/r}-\frac{(\vec{k}+\vec{p})^2}{2M}+i\eps}\equiv {} \frac{-i 2 M g^2{\cal O}_{22,\ i m A \alpha }^{\phantom{xx}\ j n B \beta}}{2 M E-r{k^2}-r{p^2}-{(\vec{k}+\vec{p})^2}+i\eps},\notag\\
	&i{\cal M}_{23,\ i m A \alpha }^{\phantom{xx}\ j B \beta} (E, \vec{k}, \vec{p}) ={} \frac{-i gg^\prime(U_j)_{im}(\tau_A\tau_B)_{\beta\alpha}}{E-\frac{k^2}{2M/r}-\frac{p^2}{2M}-\frac{(\vec{k}+\vec{p})^2}{2M/r}+i\eps} \equiv \frac{-i2M gg^\prime{\cal O}_{23,\ i m A \alpha }^{\phantom{xx}\ j B \beta}}{2M E-r{k^2}-{p^2}-r{(\vec{k}+\vec{p})^2}+i\eps},\notag\\
	&i{\cal M}_{32,\ i A \alpha }^{\phantom{xx}\ j m B \beta} (E, \vec{k}, \vec{p}) ={} \frac{-i gg^\prime(U_i)_{m j}(\tau_A\tau_B)_{\beta\alpha}}{E-\frac{k^2}{2M}-\frac{p^2}{2M/r}-\frac{(\vec{k}+\vec{p})^2}{2M/r}+i\eps}\equiv {} \frac{-i 2M gg^\prime{\cal O}_{32,\ i A \alpha }^{\phantom{xx}\ j m B \beta}}{2M E-{k^2}-r{p^2}-r{(\vec{k}+\vec{p})^2}+i\eps},\notag\\
	&i{\cal M}_{44,\ i m A \alpha }^{\phantom{xx}\ j n B \beta} (E, \vec{k}, \vec{p})={} \frac{-i g^{\prime 2}(U_i U_j)_{n m}(\tau_A\tau_B)_{\beta\alpha}}{E-\frac{k^2}{2M/r}-\frac{p^2}{2M/r}-\frac{(\vec{k}+\vec{p})^2}{2M/r}+i\eps} \equiv {} \frac{-i 2M g^{\prime 2}{\cal O}_{44,\ i m A \alpha }^{\phantom{xx}\ j n B \beta}}{2ME-r{k^2}-r{p^2}-r{(\vec{k}+\vec{p})^2}+i\eps}.
\end{align}
The operators $\mathcal{O}_{ij}$, which represent the spin--isospin part of the transition amplitude between channels $i$ and $j$, are implicitly defined by Eq.~\eqref{eq: tree_amp}.
The on-shell relation in the center-of-mass (c.m.) frame for the $i$-th dimer--particle channel reads
\begin{equation}\label{eq: onshell}
    E=\frac{k^2}{2\tilde{\mu}_i}-\frac{\gamma^{(\prime) 2}}{2\mu^{(\prime)}}=\frac{k^2}{2\tilde{\mu}_i}-E_B^{(\prime)},
\end{equation}
where $\tilde{\mu}_i$ denotes the dimer--particle reduced mass of the $i$-th $ZB$ channel,
given by
\begin{align}\label{eq: onshellmass}
    {\tilde{\mu}_1}^{-1} &= \frac{r}{M(1+r)} + \frac{1}{M},\quad {\tilde{\mu}_2}^{-1} = \frac{r}{M(1+r)} + \frac{r}{M},\notag\\
    {\tilde{\mu}_3}^{-1} &= \frac{r}{M(1+r^\prime)} + \frac{1}{M},\quad
    {\tilde{\mu}_4}^{-1} = \frac{r}{M(1+r^\prime)} + \frac{r}{M}.
\end{align}
Here, $\gamma^{(\prime)}=\sqrt{2\mu^{(\prime)}E_B^{(\prime)}}$ denotes the binding momentum of the dimer field with respect to the corresponding two bottom meson threshold, where $E_B^{(\prime)}$ is the binding energy of the $Z^{(\prime)}$ dimer relative to that threshold.

\subsection{Partial wave projection}
In reactions governed by the strong interaction, angular momentum and isospin provide a natural way to organize the quantum states, substantially reducing the complexity of the system and facilitating the interpretation of experimental data.
In general, the scattering amplitudes $T$ can be decomposed in a series of partial waves as
\begin{align}
    T(E,\vec{k},\vec{p}) = \sum_{L = 0}^{\infty} \: (2L + 1) \: T_{(L)} (E,k,p) \: P_{L}(\cos{\theta}) \:,
\end{align}
where $P_{L}$ is a Legendre polynomial,
$\theta=\sphericalangle(\vec{k},\vec{p})$ denotes the angle between the initial and final center-of-mass momenta $\vec{k}$ and
$\vec{p}$, respectively, and $k = |\vec{k}|$ and $p = |\vec{p}|$ are their corresponding magnitudes.
By projecting $T(E,\vec{k},\vec{p})$ onto a general partial wave with angular momentum $L$, the $L$-th partial-wave amplitude is obtained as
\begin{align}
    T_{(L)} (E,k,p) = \frac{1}{2} \int_{-1}^{1} d\cos \theta \: P_L (\cos \theta) \: T(E, \vec{k}, \vec{p}) \:,
\end{align}
Considering the tree-level amplitudes introduced above, the partial-wave projection of ${\cal M}_{11}$ yields
\begin{align}
    {\cal M}_{11,\ i A \alpha }^{(L)\ j B \beta}(E,k,p)&\equiv V_{11}^{(L)}(E,k,p){\cal O}_{11,\ i A \alpha }^{\phantom{xx}\ j B \beta}\notag\\
    &=\frac12\int_{-1}^1 d\cos \theta P_L (\cos \theta) {\cal M}_{11,\ i A \alpha }^{\phantom{xx}\ j B \beta}(E, \vec{k}, \vec{p})\notag\\
    &\phantom{}=\frac12\int_{-1}^1 d\cos \theta \frac{- 2 M g^2{\cal O}_{11,\ i A \alpha }^{\phantom{xx}\ j B \beta}P_L (\cos \theta)}{2 M E-(r+1){k^2}-(r+1){p^2}-2r{k p}\cos\theta+i\eps}\notag\\
    &\phantom{}=(-1)^L\frac{ M g^2}{r k p}Q_L\left(\frac{(r+1)(k^2+p^2)-2 M E}{2r k p}\right){\cal  O}_{11,\ i A \alpha }^{\phantom{xx}\ j B \beta},
\end{align}
where the logarithmic function $Q_L$ originates from the one-meson exchange contribution, whose definition is given by
\begin{align}
    Q_L(\beta-i\eps)\equiv\frac{(-1)^{L}}{2}\int_{-1}^{+1} dx \frac{P_{L} (x)}{x+(\beta-i\eps)}=\begin{cases}
        &\frac12\ln\frac{\beta+1-i\eps}{\beta-1-i\eps}\quad\text{for }L=0, \\&-\frac12\left(2-\beta\ln\frac{\beta+1-i\eps}{\beta-1-i\eps}\right)\quad\text{for }L=1,
        \\&\frac14\left(-6\beta+(-1+3\beta^2)\ln\frac{\beta+1-i\eps}{\beta-1-i\eps}\right)\quad\text{for }L=2.
    \end{cases}
\end{align}
Applying the same partial-wave projection to all dimer--particle scattering amplitudes, one obtains five independent transition potentials,
\begin{align}
    V_{11}^{(L)}(E,k,p)&=(-1)^L\frac{ M g^2}{r k p}
    Q_L\left(\frac{(r+1)(k^2+p^2)-2 M E}{2r k p}\right),\notag\\
    V_{22}^{(L)}(E,k,p)&=(-1)^L\frac{ M g^2}{ k p}
    Q_L\left(\frac{(r+1)(k^2+p^2)-2 M E}{2 k p}\right),\notag\\
    V_{23}^{(L)}(E,k,p)&=(-1)^L\frac{ M g g^\prime}{r k p}
    Q_L\left(\frac{2r k^2+(r+1)p^2-2 M E}{2r k p}\right),\notag\\
    V_{32}^{(L)}(E,k,p)&=(-1)^L\frac{ M g g^\prime}{r k p}
    Q_L\left(\frac{(r+1)k^2+2r p^2-2 M E}{2r k p}\right),\notag\\
    V_{44}^{(L)}(E,k,p)&=(-1)^L\frac{ M g^{\prime 2}}{r k p}
    Q_L\left(\frac{2r( k^2+p^2)-2 M E}{2r k p}\right).
\end{align}
The above projection provides the basic building block for the partial-wave-projected short-range production.

Now, let us concentrate on the partial wave projection of the spin and isospin structure. Following the same strategy of Ref.~\cite{Wilbring:2016bda},
we project out the desired channel by evaluating:
\begin{align}\label{eq: projection}
    T_{(L)}^{IS} \equiv \frac{1}{(2S+1)(2I+1)} \sum_{\substack{\tilde{m}\tilde{\eta}, \tilde{n}\tilde{\lambda}}} {\cal O}_{\;\tilde{n}\tilde{\lambda},\tilde{j}\tilde{\beta}}^{\dagger} \: T_{(L) \, \tilde{i}\tilde{\alpha}}^{\quad \tilde{j}\tilde{\beta}} \: {\cal O}^{\phantom{\dagger}}_{\tilde{m}\tilde{\eta},\tilde{i}\tilde{\alpha}}\:,
\end{align}
where $\tilde{i}$, $\tilde{j}$, $\tilde{m}$, and $\tilde{n}$ represent general spin indices in the given operators, while $\tilde{\alpha}$, $\tilde{\beta}$, $\tilde{\eta}$, and $\tilde{\lambda}$ denote general isospin indices. Note that for elastic scattering, the initial and final states must be identical, requiring $\tilde{\eta}=\tilde{\lambda}$ and $\tilde{m}=\tilde{n}$. 
The relevant projection operators used in this work are given below
\begin{align}\label{eq: pw_operators}
    {\cal O}_{j, i}^{S=1}(1\otimes 0\to1)=\:&\delta_{ij}\:,\quad\quad {\cal O}_{ j i}^{S=0}\left(1\otimes 1\to 0\right)=\:\frac{-1}{\sqrt{3}}\delta_{ij}\;,\notag\\
    {\cal O}_{ \ell,m n}^{S=1}\left(1\otimes 1\to 1\right)=\:&\frac{-1}{\sqrt{2}}(U_\ell)_{m n}\:,\notag\\
    {\cal O}_{\ell k, m n}^{S=2}\left(1\otimes 1\to 2\right)=\:&\frac12[\delta_{\ell m}\delta_{k n}+\delta_{\ell n}\delta_{k m}-\frac23\delta_{\ell k}\delta_{m n}]\:,\notag\\
    {\cal O}_{ \beta,A\alpha}^{I=1/2}\left(1\otimes \frac12\to \frac12\right)=\:&\frac{-1}{\sqrt{3}}(\tau_A)_{\alpha\beta}\:,\notag\\
    {\cal O}_{ j\beta,A\alpha}^{I=3/2}\left(1\otimes \frac12\to \frac32\right)=\:&\frac13[(\tau_j\tau_A)_{\alpha\beta}+\delta_{Aj}\delta_{\alpha\beta}]\:,
\end{align}
adopting the convention $(U_i)_{jk}=-i\epsilon_{ijk}$. 
By applying the projection equation in Eq.~\eqref{eq: projection} together with the operators in Eq.~\eqref{eq: pw_operators} to the tree-level amplitudes, we obtain the partial-wave coefficients, which are summarized in Table~\ref{Tab: coeff-spin} and Table~\ref{Tab: coeff-isospin} for the spin and isospin sectors, respectively.
\begin{table}[htbp]
    \centering
    \renewcommand\arraystretch{1.0}
    \caption{Coefficients of the spin projections.\label{Tab: coeff-spin}}
    \begin{tabular}{c c c c c c}
        \hline
        \hline
        Channel & ${\cal O}_{11}$ & ${\cal O}_{22}$ & ${\cal O}_{32}$ & ${\cal O}_{23}$ & ${\cal O}_{44}$ \\
        \hline
        $S=0$  & $0$  & $1$	&	$0 $ &	$0$ &	$-2$ \\
        $S=1$  & $1$  & $-1$	&	$\sqrt{2}$ &	$\sqrt{2}$ &	$1$ \\
        $S=2$  & $0$  & $1$	&	$ 0$ &	$0$ &	$1$ \\
        \hline
        \hline
    \end{tabular}
\end{table}
\begin{table}[htbp]
    \centering
    \renewcommand\arraystretch{1.0}
    \caption{Coefficients of the isospin projections.\label{Tab: coeff-isospin}}
    \begin{tabular}{c c }
        \hline
        \hline
        Channel & ${\cal O}_{ij}$  \\
        \hline
        $I=1/2$  & $-1$   \\
        $I=3/2$  & $2$ \\
        \hline
        \hline
    \end{tabular}
\end{table}

\section{The short-range production in NREFT \label{sec:ppd}}
\begin{figure*}[htbp]
    \begin{center}
        \includegraphics[width=1.0\textwidth]{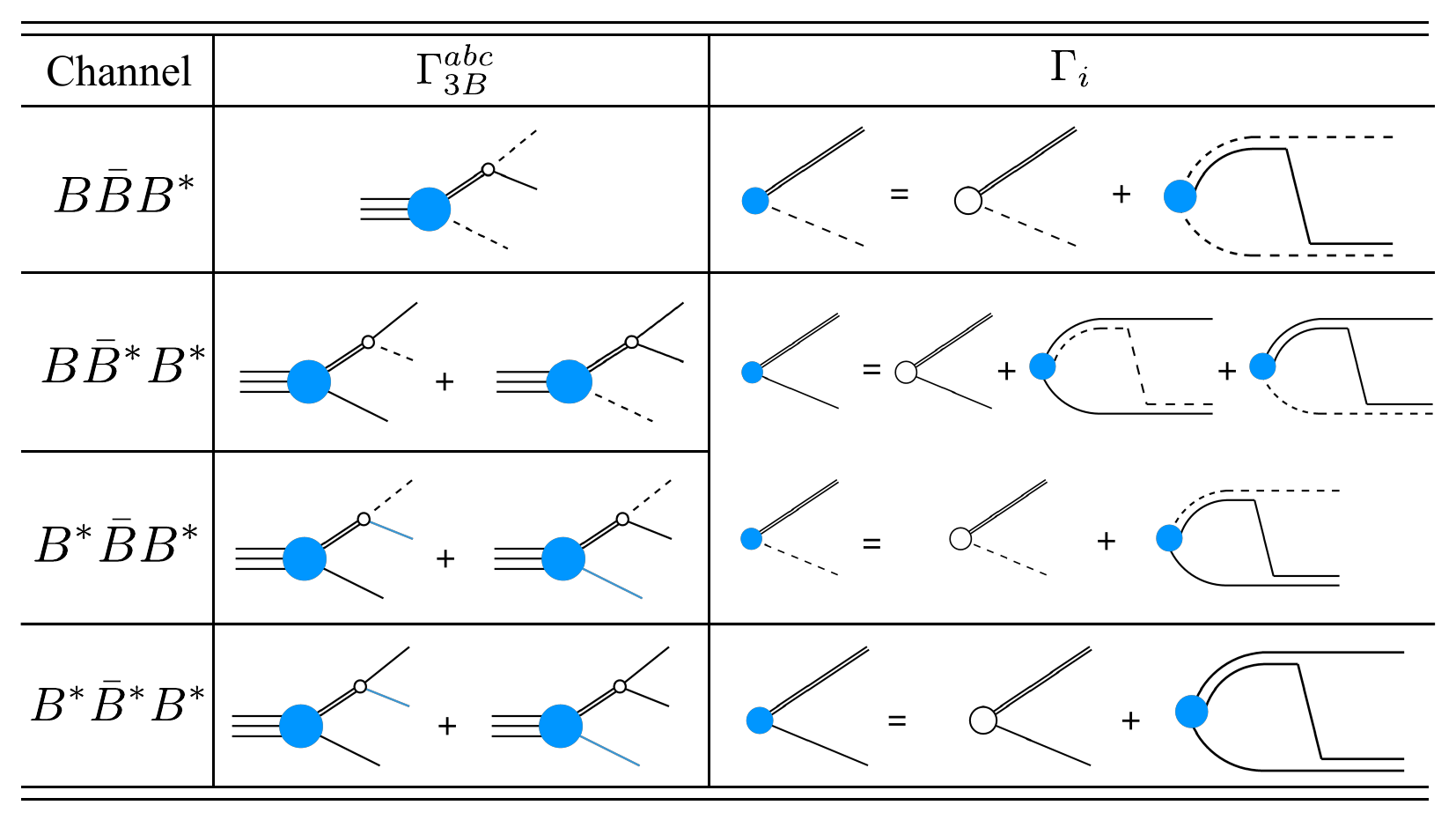}
    \end{center}
    \caption{\label{fig: pointpd}{Diagrams for the short-range productions involved in this work for the three $B$ meson systems. The second column shows the short-range production amplitudes $\Gamma_{3B}^{abc}$ for three-body systems composed of three $B$ mesons, with $abc \in $ \{$B\bar{B}B^*$, $B\bar{B}^*B^*$, $B^*\bar{B}B^*$, $B^*\bar{B}^*B^*$\}. The third column presents the short-range production amplitudes $\Gamma_i$ for the two-body dimer--particle channels, where the channel index $i=1,\ldots,4$ corresponds to the \{$ZB$, $ZB^*$, $Z^\prime B$, $Z^\prime B^*$\} channels, respectively. Note that $\Gamma_2$ and $\Gamma_3$ in the second and third rows are coupled. And the colored lines represent the symmetrization required by the presence of identical particles in the final states of the three-bottom-meson production processes. Dashed (solid) lines denote the $B$ ($B^*$) mesons, while double lines give the $Z$ and $Z^\prime$ dimers. }}
\end{figure*}
\begin{figure*}[htb]
    \begin{center}
        \includegraphics[width=0.48\textwidth]{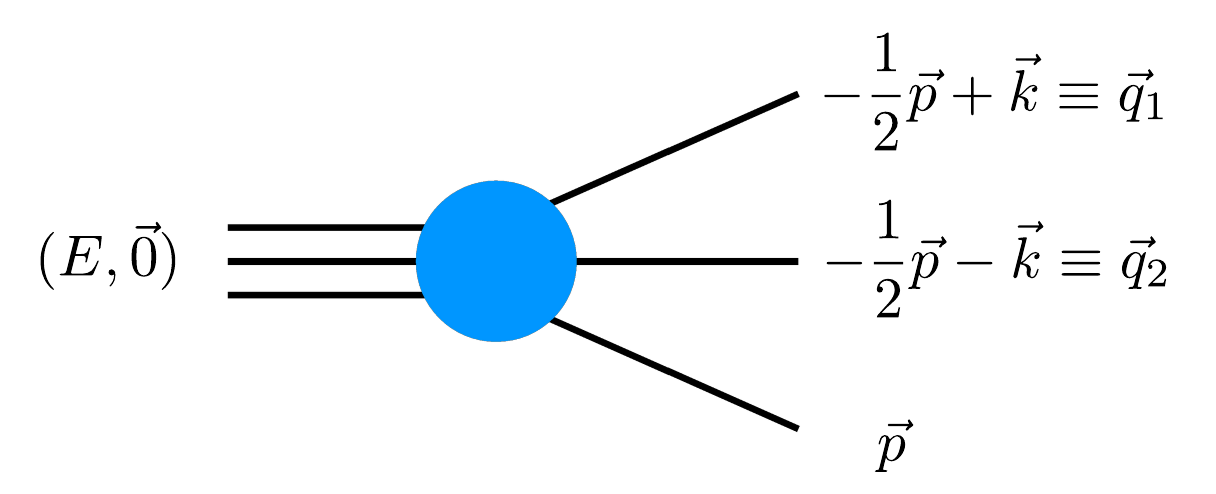}\
        \includegraphics[width=0.40\textwidth]{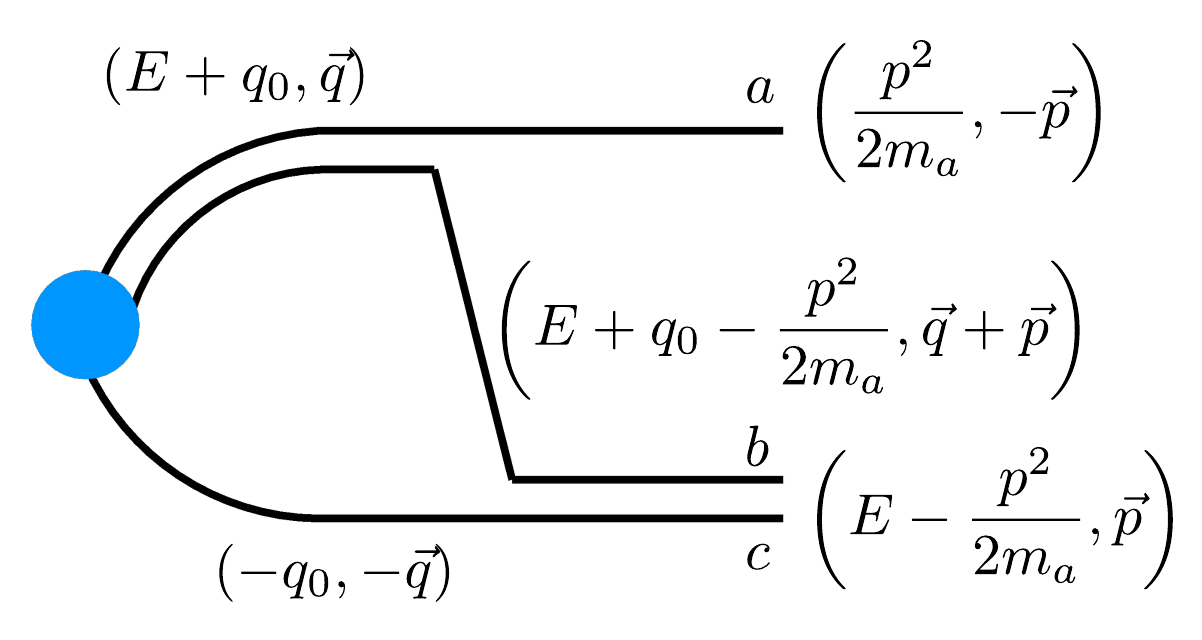}
    \end{center}
    \caption{\label{fig: kinematic-3B}{Kinematic definitions for the short-range production of three-body (left) and two-body (right) final states. The labels $a$, $b$, and $c$ denote $B$ or $B^*$ mesons.
    }}
\end{figure*}

The diagrams for the short-range production mechanisms of three-$B$-meson systems are shown in Figure~\ref{fig: pointpd}. The second column displays the short-range production amplitudes $\Gamma_{3B}^{abc}$ for the four three-body channels with $abc \in $ \{$B\bar{B}B^*$, $B\bar{B}^*B^*$, $B^*\bar{B}B^*$, $B^*\bar{B}^*B^*$\}.
The third column presents the short-range production amplitudes $\Gamma_i$ for the two-body dimer--particle channels, where the index $i=1,\ldots,4$ corresponds to the \{$ZB$, $ZB^*$, $Z^\prime B$, $Z^\prime B^*$\} channels, respectively.
In our framework, the three-meson final states are generated through all quantum-number-allowed $ZB$ intermediate channels, produced by local point sources. The short-range $ZB$ production is then described by the Skorniakov-Ter-Martirosian (STM) equation, as discussed in Refs.~\cite{Braaten:2004rn,Braaten:2021iot}.
Among these channels, the $Z B$ and $Z^\prime B^*$ systems are governed by independent single-channel STM equations, whereas the $Z B^*$ and $Z^\prime B$ channels are coupled and require a two-channel STM treatment. The kinematic definitions relevant for the short-range production processes are summarized in Figure~\ref{fig: kinematic-3B}.

\subsection{The short-range production of the two-body dimer--particle channels}
The integral equation for the short-range production amplitude $\Gamma_{1, (L)}$ of the $ZB$ two-body dimer--particle channel with definite partial wave $L$ reads 
\begin{align}\label{eq: ZB0}
    &\Gamma^{j B \beta}_{1, (L)}(E,{p}) = \sqrt{W}A_{1, (L)}^{j B \beta} (E, {p}) \notag\\
    &\phantom{xxxx}+  \int_0^\infty \frac{d q\, q^2 }{2\pi^2} \: \Gamma_{1, (L)}^{\ell C \rho}(E, {q})D\left(E-\frac{q^2}{2M},q\right)V_{11}^{(L)}(E,q,p){\cal O}_{11,\ \ell C \rho }^{\phantom{xx}\ j B \beta}.
\end{align}
Projecting on the given spin--isospin quantum number, one obtains
\begin{align}\label{eq: ZB}
    &\Gamma_{1, (L)}(E,{p}) = \sqrt{W}A_{1, (L)} (E, {p}) \notag\\
    &\phantom{xxxx}+  \int_0^\infty \frac{d q\, q^2 }{2\pi^2} \: \Gamma_{1, (L)}(E, {q})D\left(E-\frac{q^2}{2M},q\right)V_{11}^{(L)}(E,q,p)\langle{\cal O}_{11}\rangle.
\end{align}
$A_{i, (L)}$ denotes the bare amplitude for the short-range production of the $i$-th dimer--particle channel. For an $L$-wave process, it generally takes the form $g_0 p^L$, where $g_0$ is an unknown coupling constant.
$\langle {\cal O}_{11}\rangle$ represents the spin--isospin projection factor associated with the operator ${\cal O}_{11,\ \ell C \rho }^{\phantom{xx}\ j B \beta}$ in a given $(I,S)$ channel.
The corresponding coefficients are collected in Tables~\ref{Tab: coeff-spin} and \ref{Tab: coeff-isospin}. 
For instance, in the $ZB$ channel with $(I,S)=(1/2,1)$, one has $\langle {\cal O}_{11}\rangle_{IS}=1\times (-1)=-1$.

Similarly, for the short-range production amplitude $\Gamma_{4, (L)}$ of the $Z^\prime B^*$ channel, we have 
\begin{align}\label{eq: ZpBstar}
    &\Gamma_{4, (L)}(E,{p}) = \sqrt{W^\prime}A_{4, (L)} (E, {p}) \notag\\
    &\phantom{xxxx}+  \int_0^\infty \frac{d q\, q^2 }{2\pi^2} \Gamma_{4, (L)}(E, {q})D^\prime\left(E-\frac{q^2}{2M/r},q\right)V_{44}^{(L)}(E,q,p)\langle{\cal O}_{44}\rangle.
\end{align}

The coupled-channel STM equation for the short-range production amplitudes $\Gamma_{2, (L)}$ and $\Gamma_{3, (L)}$, corresponding to the$Z B^*$ and $Z^\prime B$ channels, is given by
\begin{align}\label{eq: coupledSTM}
    &\Gamma_{2, (L)}(E,{p}) = \sqrt{W}A_{2, (L)} (E, {p}) \notag\\
    &\phantom{xxxx} +  \int_0^\infty \frac{d q\, q^2 }{2\pi^2} \: \Gamma_{2, (L)}(E, {q})D\left(E-\frac{q^2}{2M/r},q\right)V_{22}^{(L)}(E,q,p)
    \langle{\cal O}_{22}\rangle\:\notag\\
    &\phantom{xxxx}+ \sqrt{\frac{W}{W^\prime}} \int_0^\infty \frac{d q\, q^2 }{2\pi^2} \: \Gamma_{3, (L)}(E, {q})D^\prime\left(E-\frac{q^2}{2M},q\right)V_{32}^{(L)}(E,q,p)\langle{\cal O}_{32}\rangle\:,\notag\\
    &\Gamma_{3, (L)}(E,{p}) = \sqrt{W^\prime} A_{3, (L)}(E, {p}) \notag\\
    &\phantom{xxxx} + \sqrt{\frac{W^\prime}{W}} \int_0^\infty \frac{d q\, q^2 }{2\pi^2} \: \Gamma_{2, (L)}(E, {q})D\left(E-\frac{q^2}{2M/r},q\right)V_{23}^{(L)}(E,q,p)\langle{\cal O}_{23}\rangle\:.
\end{align}

\subsection{The short-range production of the three-body channels}
Now, we turn to the short-range production of three $B$ mesons. As illustrated in the second column of Figure~\ref{fig: pointpd}, the short-range production of three-body final states is constructed by combining the dimer--particle production amplitude with the subsequent dimer propagator and the transition vertex connecting the dimer to its constituents.
As a result, we have
\begin{align}\label{eq: 3B}
    &\Gamma_{3B,(L)}^{B\bar{B}B^*}(E,\vec{p},\vec{k})=\Gamma_{1, (L)}(E,p)D\left(E-\frac{p^2}{2M},-\vec{p}\right)P_L(\cos\theta_{1}),\notag\\
    &\Gamma_{3B,(L)}^{B\bar{B}^*B^*}(E,\vec{p},\vec{k})=\Gamma_{2, (L)}(E,p)D\left(E-\frac{p^2}{2M/r},-\vec{p}\right)P_L(\cos\theta_{1})\notag\\
    &\phantom{xxxx}+\Gamma_{3, (L)}(E,{q_2})D^\prime\left(E-\frac{q_2^2}{2M},-\vec{q}_2\right)P_L(\cos\theta_{2}),\notag\\
    &\Gamma_{3B,(L)}^{B^*\bar{B}B^*}(E,\vec{p},\vec{k})=\Gamma_{2, (L)}(E,{p})D\left(E-\frac{p^2}{2M/r},-\vec{p}\right)P_L(\cos\theta_{1})\notag\\
    &\phantom{xxxx}+\Gamma_{2, (L)}(E,{q_2})D\left(E-\frac{q_2^2}{2M/r},-\vec{q}_2\right)P_L(\cos\theta_{2}),\notag\\
    &\Gamma_{3B,(L)}^{B^*\bar{B}^*B^*}(E,\vec{p},\vec{k})=\Gamma_{4, (L)}(E,{p})D^\prime\left(E-\frac{p^2}{2M/r},-\vec{p}\right)P_L(\cos\theta_{1)})\notag\\
    &\phantom{xxxx}+\Gamma_{4, (L)}(E,{q_2})D^\prime\left(E-\frac{q_2^2}{2M/r},-\vec{q}_2\right)P_L(\cos\theta_{2}),
\end{align}
with $\vec{q}_2=-\vec{k}-\vec{p}/2$ and $q_2 = |\vec{q}_2|$ is the corresponding magnitude. $\theta_1=\sphericalangle(\vec{p},\vec{k})$, and $\theta_2=\sphericalangle(\vec{q}_2,\vec{k})$.
$\Gamma_{i, (L)}$ on the right-hand side denotes the short-range production amplitude of the corresponding dimer--particle channel, obtained as the solution of the STM equations in Eqs.~\eqref{eq: ZB}, \eqref{eq: ZpBstar}, and \eqref{eq: coupledSTM}.
The momentum definitions refer to Figure~\ref{fig: kinematic-3B}, and one has
\begin{equation}
    \cos\theta_{2}=-\frac{(p \cos\theta_{1})/2+k}{\sqrt{p^2/4+k^2+pk\cos\theta_{1}}}.
\end{equation}

Furthermore, the symmetrized amplitude can be employed to calculate the short-range production rate function $R(E)$, which is defined as the phase-space integral of the full short-range production amplitude,
\begin{align}
    &R_{3B}^{abc}(E)=\int \frac{d^3 k}{(2\pi)^3}\int \frac{d^3 p}{(2\pi)^3}\left|\sum_{L}(2L+1)\Gamma_{3B,(L)}^{abc}(E,\vec{p},\vec{k})\right|^22\pi\delta\left(E-E^a_{\vec{p}}-E^b_{\vec{q}_1}-E^c_{\vec{q}_2}\right),
\end{align}
where the $\delta$-function ensures the energy conservation of the three outgoing mesons, which can be expressed in terms of the kinematic variables as
\begin{align}
    &\delta\left(E-E^a_{\vec{p}}-E^b_{-\frac{\vec{p}}{2}+\vec{k}}-E^c_{-\frac{\vec{p}}{2}-\vec{k}}\right)=\delta\left(E-\frac{p^2}{2m_a}-\frac{p^2/4+k^2}{2m_b m_c}\Delta^{bc}_+-\frac{pk}{2 m_b m_c} \Delta^{bc}_- z\right),
\end{align}
with $z\equiv \cos\theta_{1}$ and $\Delta_\pm^{bc}=m_b\pm m_c$. 
Here, the three-bottom-meson channel $abc$ runs over $\{B\bar{B}B^*, B\bar{B}^*B^*, B^*\bar{B}B^*, B^*\bar{B}^*B^*\}$ with $m_i$ ($i=a, b, c$) the mass of the corresponding bottom meson.  
The phase space integral can be further reduced to
\begin{equation}
    \int \frac{d^3 k}{(2\pi)^3}\int \frac{d^3 p}{(2\pi)^3}=\frac{8\pi^2}{(2\pi)^6}\int_0^\infty d k\, k^2 \int_0^\infty d p\, p^2\int_{-1}^{1} d z.
\end{equation}
Combining all ingredients, one obtains
\begin{align}\label{eq: 3Brate}
    R_{3B}^{abc}(E)&=\frac{16\pi^3}{(2\pi)^6}\int_0^\infty d k\, k^2 \int_0^\infty d p\, p^2\int_{-1}^{1} d z\left|\sum_{L}(2L+1)\Gamma_{3B,(L)}^{abc}(E,\vec{p},\vec{k})\right|^2\notag\\
    &\phantom{xxxx}\times\delta\left(E-\frac{p^2}{2m_a}-\frac{p^2/4+k^2}{2m_b m_c}\Delta^{bc}_+-\frac{pk}{2 m_b m_c} \Delta^{bc}_- z\right)\notag\\
    &= \frac{1}{4\pi^3}\int_0^\infty d k\, k^2 \int_0^\infty d p\, p^2\frac{2m_b m_c}{|pk\Delta^{bc}_-|}\left|\sum_{L}(2L+1)\Gamma_{3B,(L)}^{abc}(E,\vec{p},\vec{k})\right|_{z=z_0}^2\theta(1-|z_0|),
\end{align}
with
\begin{equation}
    z_0= \frac{2m_a m_b m_c E-m_b m_c{p^2}-m_a{(p^2/4+k^2)}\Delta^{bc}_+}{m_a pk\Delta^{bc}_-}.
\end{equation}
Similarly, one can also consider the short-range production rate for the $i$-th dimer--particle process, that is given by
\begin{align}
    &R^{i}(E)=\int_{-1}^{1}\frac{d z}{2\pi}\left( \left|\sum_L(2L+1)\Gamma_{i,(L)}(E,{p})P_L(z)\right|^2 {\tilde{\mu}_i p}\right)\bigg|_{p=\sqrt{2\tilde{\mu}_i(E+E_B^{(\prime)})}}.
\end{align}

\section{Asymptotic universality and finite-range correction \label{sec:cft}}
Before presenting the numerical results, we take a moment to discuss two complementary aspects of short-range production: asymptotic universality and finite-range correction. The former serves as a benchmark for identifying unparticle-like behavior, while the latter accounts for theoretical uncertainties arising from the finite-range effects in the effective range expansion. These corrections account for the finite-range interaction effects between two $B$ mesons, which originate from one-particle exchange dynamics.

\subsection{Asymptotic universality}
As outlined in the Introduction, the asymptotic behavior of short-range production amplitudes, dictated by conformal field theory, provides a key diagnostic for identifying whether the system exhibits the universal power-law scaling characteristics associated with unparticle dynamics.
In the unitarity limit that $a\to \infty$ and $E\to 0$, the LO integral equation in Eq.~\eqref{eq: ZB}, neglecting the inhomogeneous terms, can be rewritten as ({let $\lambda\equiv\frac{r}{1+r}=0.497862$ and $\lambda^\prime\equiv\frac{r^\prime}{1+r^\prime}=\frac12$})
\begin{align}\label{eq: ZB_asy}
    &\Gamma^{\text{asy}}_{1, (L)}(E,{p}) =\langle{\cal O}_{11}\rangle \int_0^\infty \frac{d q \,q^2}{2\pi^2}\: \Gamma_{1, (L)}^{\text{asy}}(E, {q}){ \left[\frac{\pi}{2 q p}\frac{M}{r\mu}\frac{(-1)^L
            Q_L\left(\frac{(r+1)(q^2+p^2)}{2r q p}\right) }{ \sqrt{q^2 \mu/\tilde{\mu}_1 - i\eps}}\right] }\:\notag\\
    &\phantom{xxx}= \frac{(-1)^L\langle{\cal O}_{11}\rangle}{4\pi} \int_0^\infty \frac{d q}{p}  
    {\Gamma_{1, (L)}^{\text{asy}}(E, {q})}{ \frac{1}{\lambda \sqrt{1-\lambda^2}}Q_L\left(\frac{q^2+p^2}{2\lambda q p}\right) }\:.
\end{align}
The equation satisfied by a solution exhibiting the power-law behavior, $\Gamma^{\text{asy}}_{1, (L)}(E,{p})=p^{s-1}$, is then obtained by substituting this ansatz into Eq.~\eqref{eq: ZB_asy} and evaluating the integral over $q$,
\begin{align}\label{eq: powlaw}
    \lambda\sqrt{1-\lambda^2}=\begin{cases}
        &\frac{\langle{\cal O}_{11}\rangle}{4}f(s,\lambda)\quad\text{for the $S$-wave,}\\
        &\frac{\langle{\cal O}_{11}\rangle}{4}\frac{-1}{2\lambda}\left[f(s+1,\lambda)+f(s-1,\lambda)\right]\quad\text{for the $P$-wave,}\\
        &\frac{\langle{\cal O}_{11}\rangle}{4}\frac32\frac1{4\lambda^2}\left[2f(s,\lambda)+f(s+2,\lambda)+f(s-2,\lambda)\right]\quad\text{for the $D$-wave.}
    \end{cases}
\end{align}
Here, we introduce the function $f(s,x)$ to simplify the notation,
\begin{equation}
    f(s,x)=\frac{\sin\left(s \arcsin x\right)}{s\cos\left(s\pi/2\right)}.
\end{equation}
The smallest positive solution $s$ of the asymptotic equation in Eq.~\eqref{eq: powlaw} for a given channel determines the characteristic scaling dimension $\Delta$ of the operator that creates the unparticle in the system, given by $\Delta=s+5/2$. 
Together with the total mass of the corresponding three-body system, this scaling dimension uniquely defines a nonrelativistic unparticle within the framework of nonrelativistic conformal field theory, as discussed in Refs.~\cite{Hammer:2021zxb,Braaten:2021iot}.   
The scaling dimension $\Delta$ further dictates the power-law dependence of the production rate, $R(E) \propto E^{\Delta - 3}$, which can be tested numerically to determine whether unparticle behavior emerges in the system.
For the $S$-wave $ZB$ channel with $(I,S)=({3}/2(1/2),1)$, 
we obtain $s_1=0.595214(1.18288)$, corresponding to the power-law behavior $E^{0.095214(0.68288)}$ for $(I,S)=(3/2(1/2),1)$. 
For the $P$-wave $ZB$ channel with the same quantum numbers, the solutions are $s_1=2.12862(1.94009)$, leading to a power-law behavior of $E^{1.62862(1.44009)}$.
Moreover, for the 
$D$-wave projection, we find $s_1=3.06952(2.96186)$, corresponding to $E^{2.56952(2.46186)}$.
Similarly, for the $S$-wave $Z^\prime B^*$ channel with $(I,S)=({3}/2(1/2),1)$, we obtain $s_4=0.594394(1.18266)$ and a corresponding scaling of $E^{0.094394(0.68266)}$.
For the $P$-wave $Z^\prime B^*$ channel with $(I,S)=({3}/2(1/2),1)$, the obtained value $s_4=2.12938(1.94043)$ leads to a power-law dependence of $E^{1.62938(1.44043)}$. And $s_4=3.06870(2.96144)$ with $E^{2.56870(2.46144)}$ for the $D$-wave projection. 
All extracted scaling dimensions for the $ZB$ and $Z^\prime B^*$ channels are summarized in Table~\ref{Tab: scaling_1}.
\begin{table}[htbp]
    \centering
    \renewcommand\arraystretch{1}
    \caption{The asymptotic scaling dimensions $s$ for the $ZB$ and $Z^\prime B^*$ single channels with various quantum numbers.\label{Tab: scaling_1}}
    \begin{tabular}{c|ccc }
        \hline
        \hline
        Channel  & $S$-wave & $P$-wave & $D$-wave \\
        \hline
        $ZB (I=1/2)$           & $1.18288$     & $1.94009$      & $2.96186$  \\
        $ZB (I=3/2)$           & $0.595214$    & $2.12862$      & $3.06952$ \\
        $Z^\prime B^* (I=1/2)$ & $1.18266$     & $1.94043$      & $2.96144$  \\
        $Z^\prime B^* (I=3/2)$ & $0.594394$    & $2.12938$      & $3.06870$ \\
        \hline
        \hline
    \end{tabular}
\end{table}

The coupled asymptotic equations for the two-channel $ZB^*$-$Z^\prime B$ system read
\begin{align}\label{eq: equation_ppd_asy23}
    &\Gamma^{\text{asy}}_{2, (L)}(E,{p}) = \frac{(-1)^L\langle{\cal O}_{22}\rangle}{4\pi} \int_0^\infty \frac{d q}{p} {\Gamma_{2, (L)}^{\text{asy}}(E, {q})}{ \frac{(1+r)^2}{\sqrt{r(r+2)}} {
            Q_L\left(\frac{(r+1)(q^2+p^2)}{2 q p}\right) } }\:\notag\\
    &\phantom{xxxxxxx}+ \frac{(-1)^L\langle{\cal O}_{32}\rangle}{4\pi} \frac{g {\sqrt{W}}}{ g^{\prime }{\sqrt{W^\prime}}} \int_0^\infty\frac{d q}{p} {\Gamma_{3, (L)}^{\text{asy}}(E, {q})} \frac{4\sqrt{r}}{\sqrt{r+2}} {Q_L\left(\frac{(r+1)q^2+2r p^2}{2r q p}\right)} \:,\notag\\
    &\Gamma^{\text{asy}}_{3, (L)}(E,{p}) = \frac{(-1)^L\langle{\cal O}_{23}\rangle}{4\pi}\frac{ g^{\prime }{\sqrt{W^\prime}}}{g {\sqrt{W}}} \int_0^\infty \frac{d q}{p} { {\Gamma_{2, (L)}^{\text{asy}}(E, {q})\frac{(1+r)^2}{r\sqrt{r(r+2)}}Q_L\left(\frac{2r q^2+(r+1)p^2}{2r q p}\right)} }\:.
\end{align}
{Note that in the coupled-channel case, rather than solving the equations directly for $s_2$ and $s_3$, we follow the strategy of Ref.~\cite{Hildenbrand:2019sgp} by introducing a transformation of $\Gamma_2$ and $\Gamma_3$ into two new amplitudes that decouple the integral equations.
To absorb the dependence on kinematic variables into the amplitudes, we replace the normalization used in Eq.~\eqref{eq: equation_ppd_asy23}, $\Gamma_{2,(L)}=\sqrt{W} \Gamma_{2,(L)}^{\rm bare}$ and $\Gamma_{3,(L)}=\sqrt{W^\prime} \Gamma_{3,(L)}^{\rm bare}$, with
$\Gamma_{2, (L)}=\sqrt{W} \Gamma_{2,(L)}^{\rm bare}$ and $\Gamma_{3,(L)}=\frac{g}{g^\prime}\sqrt{W}\Gamma_{3, (L)}^{\rm bare}$.
Rewriting the equations in matrix form, we obtain: 
\begin{equation}\label{eq: ZBstar-ZpB}
    \begin{pmatrix}
        \Gamma^{\text{asy}}_{2, (L)} \\
        \Gamma^{\text{asy}}_{3, (L)}
    \end{pmatrix}=\int_0^\infty \frac{d q (-1)^L}{4\pi p}\begin{pmatrix}
        \frac{(1+r)^2 \langle{\cal O}_{22}\rangle}{\sqrt{r(r+2)}} {
            Q_L\left(\frac{(r+1)(q^2+p^2)}{2 q p}\right) } & \frac{4\sqrt{r} \langle{\cal O}_{32}\rangle}{\sqrt{r+2}} {Q_L\left(\frac{(r+1)q^2+2r p^2}{2r q p}\right)}  \\
        \frac{(1+r)^2\langle{\cal O}_{23}\rangle}{r\sqrt{r(r+2)}}Q_L\left(\frac{2r q^2+(r+1)p^2}{2r q p}\right)  & 0
    \end{pmatrix}\begin{pmatrix}
        \Gamma_{2, (L)}^{\text{asy}} \\
        \Gamma_{3, (L)}^{\text{asy}}
    \end{pmatrix}.
\end{equation} 
Assuming the existence of a transformation that decouples the above integral equations, we express it in the general form
\begin{equation}
    \begin{pmatrix}
        \Gamma^{\text{asy}}_2(E,p)  \\
        \Gamma^{\text{asy}}_3(E,p)
    \end{pmatrix}=\begin{pmatrix}
        C_{a}^2 & C_{b}^2  \\
        C_{a}^3 & C_{b}^3 
    \end{pmatrix}\begin{pmatrix}
        \Gamma^{\text{asy}}_a(E,p)  \\
        \Gamma^{\text{asy}}_b(E,p)
    \end{pmatrix},
\end{equation} 
where the new amplitudes $\Gamma^{\text{asy}}_a(E,p)$ and $\Gamma^{\text{asy}}_b(E,p)$
satisfy two independent integral equations.
The corresponding equations for the transformation coefficients $(C^2_i, C^3_i)$ are then obtained for $i\in \{a,b\}$.
Evaluating the $q$ integral with the ansatz $\Gamma_i(E,p)=p^{s_i-1}$ for the $i$-th transformed amplitude and replacing $r$ by $\lambda$ yields
\begin{equation}\label{eq: ZBstar-ZpB_decouple}
    \begin{pmatrix}
        C^2_i  \\
        C^3_i
    \end{pmatrix}=\begin{pmatrix}
        \frac{\langle{\cal O}_{22}\rangle}{ 4}\frac{f(s_i,1-\lambda)}{(1-\lambda)\sqrt{1-(1-\lambda)^2}} & \frac{\langle{\cal O}_{32}\rangle}{ 4}\frac{4 \lambda (\sqrt{2 \lambda})^{s_i} f(s_i,\sqrt{\lambda/2}) }{\sqrt{1-(1-\lambda)^2}}  \\
        \frac{\langle{\cal O}_{23}\rangle}{4} \frac{(1/\sqrt{2 \lambda})^{s_i} f(s_i,1/(2\sqrt{2\lambda}))}{\lambda\sqrt{1-(1-\lambda)^2}}  & 0
    \end{pmatrix}\begin{pmatrix}
        C^2_i  \\
        C^3_i
    \end{pmatrix}\equiv U\begin{pmatrix}
        C^2_i  \\
        C^3_i
    \end{pmatrix},
\end{equation}
for the $S$-wave case. $s_a$ and $s_b$ satisfy this same equation. 
{The above equation has only nontrivial solutions if the determinant of the homogeneous coefficient matrix of vector $\{C_i^2,C_i^3 \}$ vanishes (i.e., ${\rm det}(A)=0$ for $A \vec{x}=0$ )}, which leads to
\begin{equation}
    {\rm det}(U-I)=1-\frac{\langle{\cal O}_{22}\rangle f(s_i,1-\lambda)}{4(1-\lambda)\sqrt{1-(1-\lambda)^2}}
    -\frac{\langle{\cal O}_{32}\rangle\langle{\cal O}_{23}\rangle
        f(s_i,\frac1{2\sqrt{2\lambda}}) f(s_i,\frac{\sqrt{\lambda}}{\sqrt{2}})}{4(1-(1-\lambda)^2)}=0.
\end{equation} 
This yields $s=0.593750(0.593657)$ for $(I,S)=({3}/2(1/2),1)$. 
For the $P$-wave case, the vanishing determinant condition of the coefficient matrix becomes
\begin{align}
    &{\rm det}(U-I)=1+\frac{\langle{\cal O}_{22}\rangle}{8(\lambda-1)^2\sqrt{1-(1-\lambda)^2}}\bigg[f(s_i+1,1-\lambda)+f(s_i-1,1-\lambda)\bigg]\notag\\
    &\phantom{xx}+\frac{\langle{\cal O}_{23}\rangle\langle{\cal O}_{32}\rangle}{8(\lambda-2)\lambda^2}\bigg[f(s_i+1,\frac1{2\sqrt{2\lambda}})f(s_i+1,\frac{\sqrt{\lambda}}{\sqrt{2}})+f(s_i-1,\frac1{2\sqrt{2\lambda}})f(s_i-1,\frac{\sqrt{\lambda}}{\sqrt{2}})\bigg]\notag\\
    &\phantom{xx}+\frac{\langle{\cal O}_{23}\rangle\langle{\cal O}_{32}\rangle}{8(\lambda-2)\lambda^2}\bigg[f(s_i+1,\frac1{2\sqrt{2\lambda}})f(s_i-1,\frac{\sqrt{\lambda}}{\sqrt{2}})+f(s_i-1,\frac1{2\sqrt{2\lambda}})f(s_i+1,\frac{\sqrt{\lambda}}{\sqrt{2}})
    \bigg]=0,
\end{align}
resulting in $s=1.77179(1.93993)$ for $(I,S)=({3}/2(1/2),1)$. Similarly, for the 
$D$-wave projection, one obtains $s=2.82599(2.96142)$ for the same $(I,S)=({3}/2(1/2),1)$ channel. 
All extracted scaling dimensions for the coupled $ZB^*$ and $Z^\prime B$ channels are summarized in Table~\ref{Tab: scaling_2}.

These scaling dimensions will be verified in the next section when we present our numerical results. Furthermore, the short-range production rates of three-body channels, obtained from the amplitudes in Eq.~\eqref{eq: 3B}, are expected to follow power-law behavior in the asymptotic region and possess the same scaling dimension as the associated two-body dimer--particle channels. 
The relation between the power-law exponent and the scaling dimension, however, differs slightly: for three-body production, the rate scales as $R(E) \propto E^{\Delta - 5/2}$.
\begin{table}[htbp]
    \centering
    \renewcommand\arraystretch{1}
    \caption{The asymptotic scaling dimensions $s$ for the coupled $ZB^*$ and $Z^\prime B$ channels with various quantum numbers. Note that all coupled channels share the same scaling dimension. \label{Tab: scaling_2}}
    \begin{tabular}{c|ccc }
        \hline
        \hline
        Channel & $S$-wave & $P$-wave & $D$-wave \\
        \hline
        $ZB^*/Z^\prime B (I=1/2)$   & $0.593657$    & $1.93993$      & $2.96142$  \\
        $ZB^*/Z^\prime B (I=3/2)$   & $0.593750$    & $1.77179$      & $2.82599$ \\
        \hline
        \hline
    \end{tabular}
\end{table}

\subsection{Finite-range correction}
Recalling the dressed dimer propagator in Eq.~\eqref{eq: propagators}, the zero-range ERE matching introduces only the scattering length $a=1/\gamma$, which absorbs the artificial scale dependence arising in the renormalization of the self-energy diagrams.
However, when the effective range expansion of the elastic scattering amplitude of the corresponding two constituent particles is carried out to linear order in the effective range $\rho$, namely,
\begin{equation}
    {k\cot\delta_0}(k)=-\gamma+\frac12 \rho\left(k^2+\gamma^2\right),
\end{equation}
an additional parameter, the effective range $\rho$, enters the description, as discussed in previous works~\cite{Phillips:1999hh,Hammer:2001gh,Bedaque:2002yg,Afnan:2003bs,Braaten:2004rn,Ebert:2021epn}. 
For later convenience, we introduce the dimensionless parameter given by the product of the effective range and the binding momentum, $\rho\gamma$. 
The effective range $\rho$ enters the integral equations in Eqs.~\eqref{eq: ZB}, \eqref{eq: ZpBstar}, and \eqref{eq: coupledSTM} through the dimer propagator, which now takes the form
\begin{align}\label{eq: dimer_NLO}
    i \tilde{D}_\rho (p_0, \vec{p}) \equiv i \tilde{D}_\rho (k^{*}) &= -i \frac{\pi}{2 g^{2} \mu} \frac1{-\gamma+ k^{*}+\frac\rho{2}(\gamma^{2}-k^{*2})}\:=-i \frac{\pi}{2 g^{2} \mu} \frac{-2/\rho}{(k^{*}-k_1)(k^{*}-k_2)}
\end{align}
with
\begin{equation}
    k^{*}= \sqrt{\frac{r}{(1+r)^2}{p}^2  -2\mu p_0 }.
\end{equation}
Note that the prescription $k^*\to k^* -i \eps$ is implicit in above expression. 
The dimer propagator thus exhibits two apparent poles,
\begin{equation}
    k_1=\gamma,\quad k_2=\frac{2}{\rho}-\gamma,
\end{equation}
which lead to two singularities on the integration contour of the loop momentum $q$ for positive $\rho$ when solving the integral equations in Eqs.~\eqref{eq: ZB}, \eqref{eq: ZpBstar} and \eqref{eq: coupledSTM} for the short-range production of those two-body dimer--particle channels. 
The corresponding expression for the $Z^\prime$ dimer is obtained straightforwardly by replacing the kinematic variables in the above equations.

Considering the finite-range correction to Eq.~\eqref{eq: ZB} for the short range production of $ZB$ channel as an explicit example, where ${D} (q_0, \vec{q}) \to \tilde{D}_\rho (q_0, \vec{q})$ with $q_0=E-q^2/M$, the two singularities are given by
\begin{equation}
    q_{\rm phy.}=\sqrt{2 \tilde{\mu}_1 (E+E_B)},\quad q_{\rm unphy.}=\sqrt{2\tilde{\mu}_1\left[E+E_B\left(1-\frac2{\rho\gamma}\right)^2\right]}.
\end{equation}
For a given energy $E$, these singularities are fixed by the binding energy $E_B$ and the dimensionless parameter $\rho\gamma$.
The first singularity clearly corresponds to the physical pole, which appears when the intermediate state goes on-shell (see Eq.~\eqref{eq: onshell}), while the second is an unphysical pole that leads to a negative residue in the corresponding two-body elastic amplitude. A more detailed discussion of this unphysical singularity can be found in Ref.~\cite{Ebert:2021epn}.

A major challenge in incorporating the finite-range effects is that the range-corrected dimer propagator introduces an unphysical singularity that lies on the integration contour for positive effective range $\rho$. 
In the literature, such spurious poles are typically handled within a strict perturbative framework, in which the two-body amplitude is expanded in the range parameter(s), and finite-range corrections are included order by order according to a well-defined power-counting scheme~\cite{Hammer:2001gh,Bedaque:1998km,Ji:2011qg,Ji:2012nj,Vanasse:2013sda}. 
An alternative approach, designed for finite-volume calculations, was proposed more recently in Ref.~\cite{Ebert:2021epn}. 
It should be emphasized that, in general, a three-body force is required to renormalize the ultraviolet divergence associated with finite-range corrections~\cite{Braaten:2004rn,Ebert:2021epn}. 
In this procedure, some known observables of the corresponding three-body system, such as the dimer--particle scattering phase shift or a three-body binding energy, is needed to fine tune the strength of the three-body force. 
However, the limited experimental and theoretical knowledge of the three-bottom-meson systems considered here prevents a fully controlled perturbative treatment of the spurious singularity. 

Instead, we solve the range-corrected integral equations by adopting a momentum cutoff that is sufficiently low to exclude the unphysical pole from the integration contour. This strategy inevitably introduces cutoff dependence into the calculation. 
To assess the robustness of this low-cutoff treatment and to quantify the associated cutoff dependence, we compare it with the expansion prescription proposed in Ref.~\cite{Ebert:2021epn}.

As shown in Ref.~\cite{Ebert:2021epn}, the range-corrected dimer propagator can be decomposed into physical and unphysical contributions.
The unphysical term can then be expanded in powers of $k^{*2}/k_2^2$, yielding
\begin{align}
    &\tilde{D}_\rho (p_0, \vec{p})=- \frac{\pi}{2 g^{2} \mu} \frac{-2/\rho}{(k^{*}-k_1)(k^{*}-k_2)}\notag\\
    &=- \frac{\pi}{2 g^{2} \mu}\bigg[\frac{2(k_1+k_2)/\rho}{(k_2-k_1)(k^*+k_2)(k^*-k_1)}-\frac{4k_2/\rho}{(k_2-k_1)(k^{*2}-k_2^2)}\bigg]\notag\\
    &=- \frac{\pi}{2 g^{2} \mu}\bigg[\frac{2(k_1+k_2)/\rho}{(k_2-k_1)(k^*+k_2)(k^*-k_1)}+\frac{4k_2/\rho}{(k_2-k_1)k_2^2}\bigg(1+\frac{k^{*2}}{k_2^2}+\frac{k^{*4}}{k_2^4}+\cdots\bigg)\bigg].
\end{align}
After expanding in the small count of $k^{*2}/k_2^2$, the resulting propagator is free of the spurious pole, at the cost of an explicit violation of unitarity. 
Nevertheless, this violation is parametrically suppressed in the physically relevant low-momentum region compared to the large scale set by $k_2$, and can be systematically reduced by including higher-order terms in the expansion~\cite{Ebert:2021epn}. 

In this subsection, we adopt this expansion prescription as a consistency check on the low-cutoff treatment within an appropriate cutoff window. 
Specifically, we truncate the expansion at first order in $k^{*2}/k_2^2$, which leads to two spurious-pole-free approximations of the dimer propagator, denoted as $\tilde{D}_\rho^{* 1}$ and $\tilde{D}_\rho^{* 2}$,
\begin{align}
    &\tilde{D}_\rho^{* 1}(p_0, \vec{p})=- \frac{\pi}{2 g^{2} \mu}\bigg(\frac{2(k_1+k_2)/\rho}{(k_2-k_1)(k^*+k_2)(k^*-k_1)}+\frac{4k_2/\rho}{(k_2-k_1)k_2^2}\bigg),\notag\\
    &\tilde{D}_\rho^{* 2}(p_0, \vec{p})=- \frac{\pi}{2 g^{2} \mu}\bigg[\frac{2(k_1+k_2)/\rho}{(k_2-k_1)(k^*+k_2)(k^*-k_1)}+\frac{4k_2/\rho}{(k_2-k_1)k_2^2}\bigg(1+\frac{k^{*2}}{k_2^2}\bigg)\bigg].
\end{align}
The difference between results obtained using the leading-order (LO) propagator $\tilde{D}_\rho^{* 1}$ and the next-to-leading-order (NLO) propagator $\tilde{D}_\rho^{* 2}$ is expected to become significant only when the unphysical pole starts to play a role. 
In such cases, a cutoff-dependent three-body force must be introduced to tame the higher-order contributions in a perturbative manner.
Our goal is therefore to identify a cutoff window in which the low-cutoff treatment, the LO expansion $\tilde{D}_\rho^{* 1}$, and the NLO expansion $\tilde{D}_\rho^{* 2}$ yield consistent results, and where the residual cutoff dependence is negligible compared to the variation induced by the finite effective range.

\begin{figure*}[htb]
    \begin{center}
        \includegraphics[width=1\textwidth]{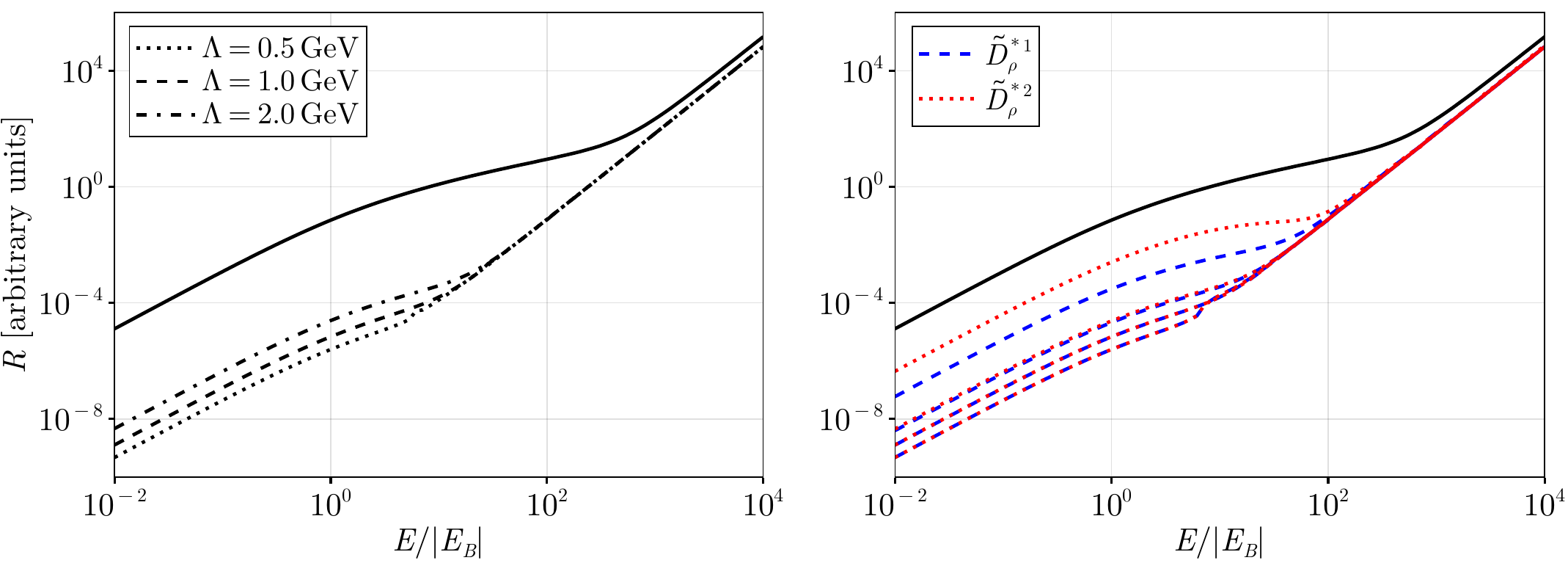}
    \end{center}
    \caption{\label{fig: ZB_erc_cutoff} Cutoff dependence of the finite-range corrections to the short-range production rates of the $B\bar{B}B^*$ channel with $(I,S)=(3/2,1)$. The left panel shows results obtained using the low-cutoff treatment of the spurious singularity for three cutoff values: $\Lambda=0.5$~GeV (dotted), $1.0$~GeV (dashed), and $2.0$~GeV (dash-dotted). The right panel displays the LO (blue dashed) and NLO (red dotted) expansion prescriptions for four cutoff values, $\Lambda=5.0$, $2.0$, $1.0$, and $0.5$~GeV, from top to bottom. The zero-range result is shown by the black solid curve for reference.
    The short-range production rates are given in arbitrary units because of the unfixed normalization parameter $g_0$ in the bare amplitude $A_{i, (L)}$ for the $i$-th dimer--particle channel; for numerical illustration, we set $g_0=1$ hereafter.
    }
\end{figure*}
Figure~\ref{fig: ZB_erc_cutoff} presents a numerical study of the cutoff dependence. 
Note that, within our framework, an unknown constant $g_0$ appears in the bare short-range production of the two-body dimer--particle channels.
This parameter introduces an arbitrary overall normalization in the magnitude of the resulting short-range production rates. 
Since $g_0$ enters only as a constant in the inhomogeneous term, it does not affect the lineshape, which is the primary focus of this work. We therefore set $g_0=1$ throughout our numerical calculations, and present all short-range production rates in arbitrary units. 
In the left panel, three low-cutoff calculations with $\Lambda=0.5$~GeV (dotted), $1.0$~GeV (dashed), and $2.0$~GeV (dash-dotted) are compared with the zero-range result (solid black).
At energies of the order of the binding energy, the variation in the short-range production rate when increasing the cutoff from $1.0$~GeV to $2.0$~GeV amounts to less than $1$\textperthousand\ of the difference between the zero-range result and the range-corrected value with $\rho\gamma=0.1$ and $\Lambda=1.0$~GeV, and is therefore numerically negligible.
 In the right panel, results obtained with the LO and NLO expanded propagators are shown for cutoff values $\Lambda=5.0$, $2.0$, $1.0$, and $0.5$~GeV, from top to bottom.
The lowest unphysical pole within the energy range considered is located at $q_{\rm unphy.}\simeq 3.5$~GeV for the short-range production of the $ZB$ channel. When the cutoff approaches or exceeds this scale, significant differences between $\tilde{D}_\rho^{* 1}$ and $\tilde{D}_\rho^{* 2}$ emerge. 
In contrast, for cutoff values in the range $0.5$-$2.0$~GeV, the two expansions yield indistinguishable results and are consistent with the corresponding low-cutoff treatment.

These results demonstrate that, for sufficiently small effective range, there exists a narrow cutoff window in which the low-cutoff strategy--keeping the cutoff below the unphysical pole--provides a numerically simple and reliable estimate of finite-range corrections. In the following calculations, we therefore fix the cutoff at $\Lambda=1.0$~GeV.

\section{Results\label{sec:results}}
We first consider the short-range production rates of two-body dimer--particle systems and investigate their asymptotic behaviors. To remain consistent with Ref.~\cite{Lin:2017dbo}, we take $E_B=5\,{\rm MeV}$ and $E_B^{\prime}=1\,{\rm MeV}$ for the illustration.
The calculated production rates for the single-channel cases ($ZB$ and $Z^\prime B^*$) and the coupled-channel case ($Z B^*$ and $Z^\prime B$) are shown in Figure~\ref{fig: rateZB_single} and Figure~\ref{fig: rateZB_coupled}, respectively.
\begin{figure*}[htb]
    \begin{center}
        \includegraphics[width=1\textwidth]{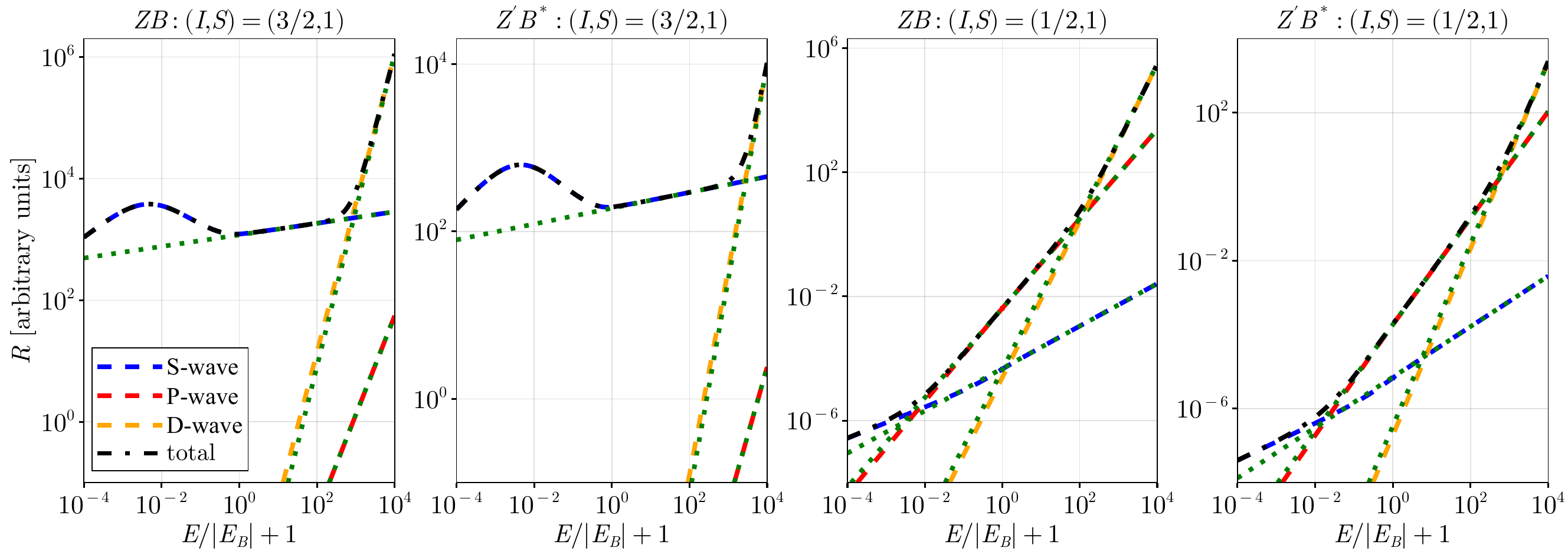}
    \end{center}
    \caption{\label{fig: rateZB_single}Short-range production rates for the $ZB$ and $Z^\prime B^*$ dimer--particle channels with $(I,S)=(3/2(1/2),1)$. The blue-dashed, red-dashed, orange-dashed and black-dash-dotted lines represent the $S$-, $P$-, $D$-wave and the total contributions, respectively. The green dotted lines show the corresponding asymptotic behavior with the scaling dimensions derived from the asymptotic equations, that is, $E^{s_i-1/2}$ with $s_i$ given by Table~\ref{Tab: scaling_1}.
    }
\end{figure*}
\begin{figure*}[htb]
    \begin{center}
        \includegraphics[width=1\textwidth]{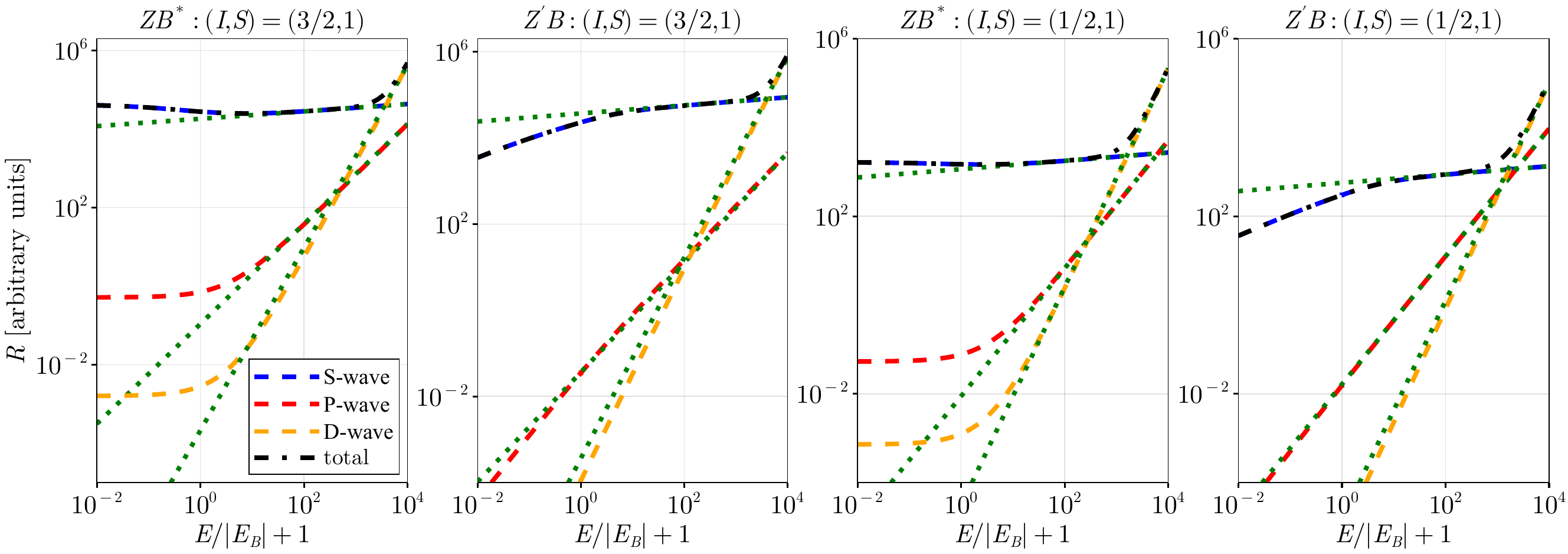}
    \end{center}
    \caption{\label{fig: rateZB_coupled}{Short-range production rates for the $ZB^*$ and $Z^\prime B$ dimer--particle channels with $(I,S)=(3/2(1/2),1)$. The notations follow those in Figure~\ref{fig: rateZB_single}, and the power-law exponents $s_i$ of $E^{s_i-1/2}$ for the green dotted lines are listed in Table~\ref{Tab: scaling_2}. 
    }}
\end{figure*}

For each channel, we compare the $S$-, $P$- and $D$-wave contributions over a broad energy range. As expected, the $S$-wave dominates near threshold, while the $P$- and $D$-wave components govern the higher-energy region, with the $D$-wave eventually overtaking the $P$-wave in the asymptotic limit. 
This results in a universal $D$-wave power-law behavior in the total rate in the high-energy limit. 
The predicted asymptotic power-law trends are accurately reproduced in all channels, and the power-law exponents of each partial wave are exactly consistent with values given in Table~\ref{Tab: scaling_1} and Table~\ref{Tab: scaling_2}, which agree well also with the corresponding predictions from the nonrelativistic conformal field theory, as demonstrated in Ref.~\cite{Braaten:2021iot}.
The power-law behavior observed in the scaling region indicates the emergence of unparticle dynamics in the short-range production processes of all $ZB$ channels with quantum number $(I,S)=(3/2(1/2),1)$, similar to the three-charm-meson unparticles investigated in Ref.~\cite{Braaten:2021iot}.
This behavior originates from the large scattering lengths that persist in the corresponding dimer--particle interactions, as investigated in Ref.~\cite{Lin:2017dbo}. 
Therefore, an experimental observation of such power-law asymptotic behavior in the short-range production rates under study would provide strong evidence for the large scattering lengths in the two $B$-meson interaction, offering a natural explanation for the dominant $B^*\bar{B}^*$ and $(B\bar{B}^*+\bar{B}B^*)/\sqrt{2}$ components in the internal structures of the $Z^\prime$ and $Z$ state, respectively.

In particular, as shown for the $I=3/2$ case in Figure~\ref{fig: rateZB_single}, the production rate $R$ increases from zero at the threshold ($E=-E_B^{(\prime)}$) to a peak near $E_{i}$ ($E_{i}=1/(2\tilde{\mu}_{i} a_{i}^2)$, giving $E_{1}=28.2\,{\rm keV}$ or $E_{4}=5.03\,{\rm keV}$ using the dimer--particle scattering lengths $a_i$ obtained in Ref.~\cite{Lin:2017dbo}), and then decreases to a local minimum at an energy of order $E_B^{(\prime)}$. Beyond the minimum lies the scaling region, where the short-range production rate follows a power-law increase. 
Moreover, as discussed in Ref.~\cite{Lin:2025ksg}, this local minimum serves as a characteristic feature that provides an experimentally accessible observable, enabling a model-independent determination of the binding energy $E_B$ of the dimer state. 
It is worth noting that the presence of such a local minimum feature depends on the quantum numbers of the channel under consideration. In our case, this feature is absent in the short-range production rates of the $(I,S)=(1/2,1)$ $ZB$ and $Z^\prime B^*$ channels, see Ref.~\cite{Lin:2025ksg} for details.

Next, we turn to the short-range production of the three-$B$-meson channels. Figures~\ref{fig: rate3B_single} and \ref{fig: rate3B_coupled} present the three-$B$-meson production rates, given by Eq.~\eqref{eq: 3Brate} and Eq.~\eqref{eq: 3B}, for the single-channel cases ($B\bar{B}B^*$ and $B^*\bar{B}^*B^*$) and the coupled-channel case ($B\bar{B}^* B^*$ and $B^*\bar{B} B^*$), respectively.
Similar to the dimer--particle case, the $S$-wave provides the dominant contribution near threshold, while the $D$-wave partial wave becomes leading in the high-energy limit. In the intermediate region, the $S$-, $P$- and $D$-wave contributions compete and coherently combine to form the total production rate.
Again, the short-range production rates of all three-$B$-meson channels exhibit clear power-law behavior in the scaling region, which is in good agreement with the corresponding prediction given in Table~\ref{Tab: scaling_1} and Table~\ref{Tab: scaling_2}, via the expression of $E^{s_i}$. 
The experimental evidence of such power-law asymptotic behavior in the three-$B$-meson short-range production rates would also provide valuable insights into the underlying two-$B$-meson interactions and the natures of $Z$ and $Z^\prime$ states. 
\begin{figure*}[htb]
    \begin{center}
        \includegraphics[width=1\textwidth]{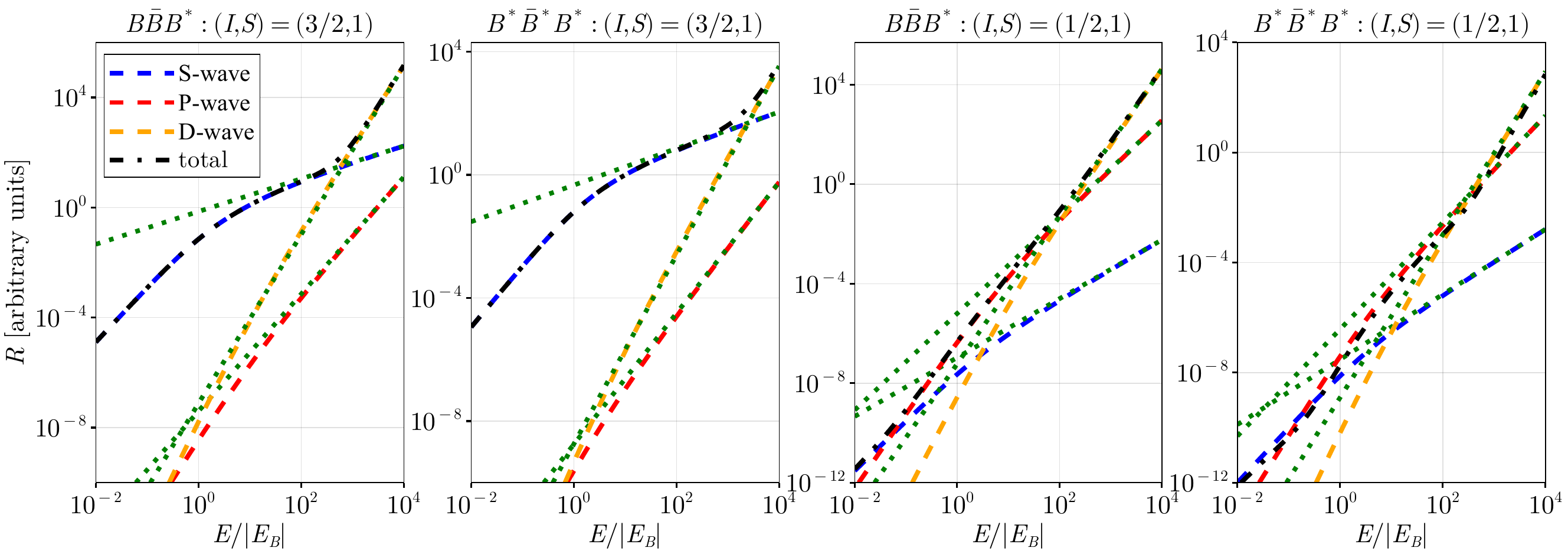}
    \end{center}
    \caption{\label{fig: rate3B_single}{Short-range production rates for the $B\bar{B}B^*$ and $B^*\bar{B}^*B^*$ three-body channels. The notations follow those in Figure~\ref{fig: rateZB_single}, and the power-law exponents $s_i$ of $E^{s_i}$ for the green dotted lines are listed in Table~\ref{Tab: scaling_1}. 
    }}
\end{figure*}
\begin{figure*}[htb]
    \begin{center}
        \includegraphics[width=1\textwidth]{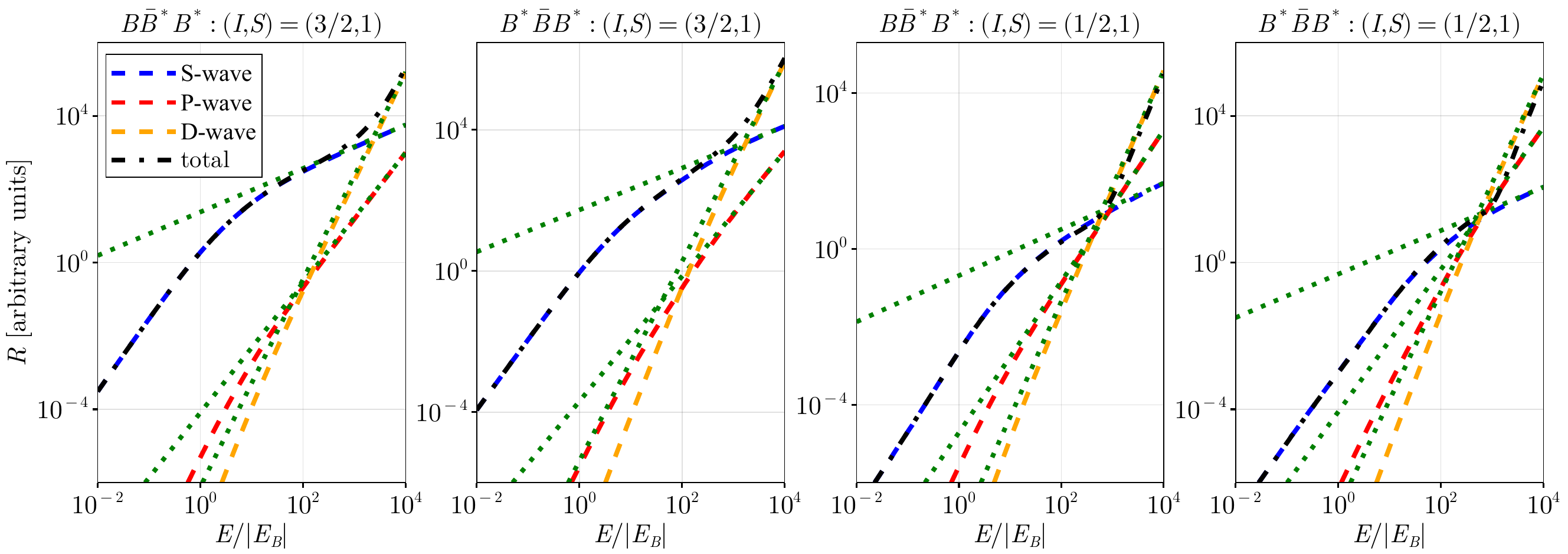}
    \end{center}
    \caption{\label{fig: rate3B_coupled}{Short-range production rates for the $B\bar{B}^*B^*$ and $B^*\bar{B}B^*$ three-body channels. The notations follow those in Figure~\ref{fig: rateZB_single}, and the power-law exponents $s_i$ of $E^{s_i}$ for the green dotted lines are listed in Table~\ref{Tab: scaling_2}. 
    }}
\end{figure*}

The finite-range corrections to the three-$B$-meson production rates are shown in Figure~\ref{fig: rate3B_single_frc} and Figure~\ref{fig: rate3B_coupled_frc} for the single-channel cases ($B\bar{B}B^*$ and $B^*\bar{B}^*B^*$) and the coupled-channel case ($B\bar{B}^* B^*$ and $B^*\bar{B} B^*$), respectively. 
As discussed in Ref.~\cite{Lin:2025ksg}, within the framework of short-range effective field theory, the short-range production rates can be reliably calculated only within the range $\rho\gamma \lesssim 0.3$. 
Extending to larger positive $\rho\gamma$ values requires introducing a nonzero three-body force to ensure ultraviolet-independent evaluation~\cite{Braaten:2004rn,Ebert:2021epn}, which is beyond the scope of the present study.
Accordingly, in Figures~\ref{fig: rate3B_single_frc} and \ref{fig: rate3B_coupled_frc}, we compare only the finite-range results with $\rho\gamma=\pm0.1$ to the zero-range calculations.
The results show that finite-range effects induce a significant shift in the three-$B$-meson production rates at low energies across all channels, while their impact becomes negligible in the high-energy region, leaving the power-law asymptotic behavior unchanged.
The sign of the finite-range parameter $\rho\gamma$ has only a moderate influence on the line shape of the three-$B$-meson short-range production rate.
The strong sensitivity of these rates to the finite-range effects provides a promising probe of the effective range parameters in the corresponding two-body subsystems.
\begin{figure*}[htb]
    \begin{center}
        \includegraphics[width=1\textwidth]{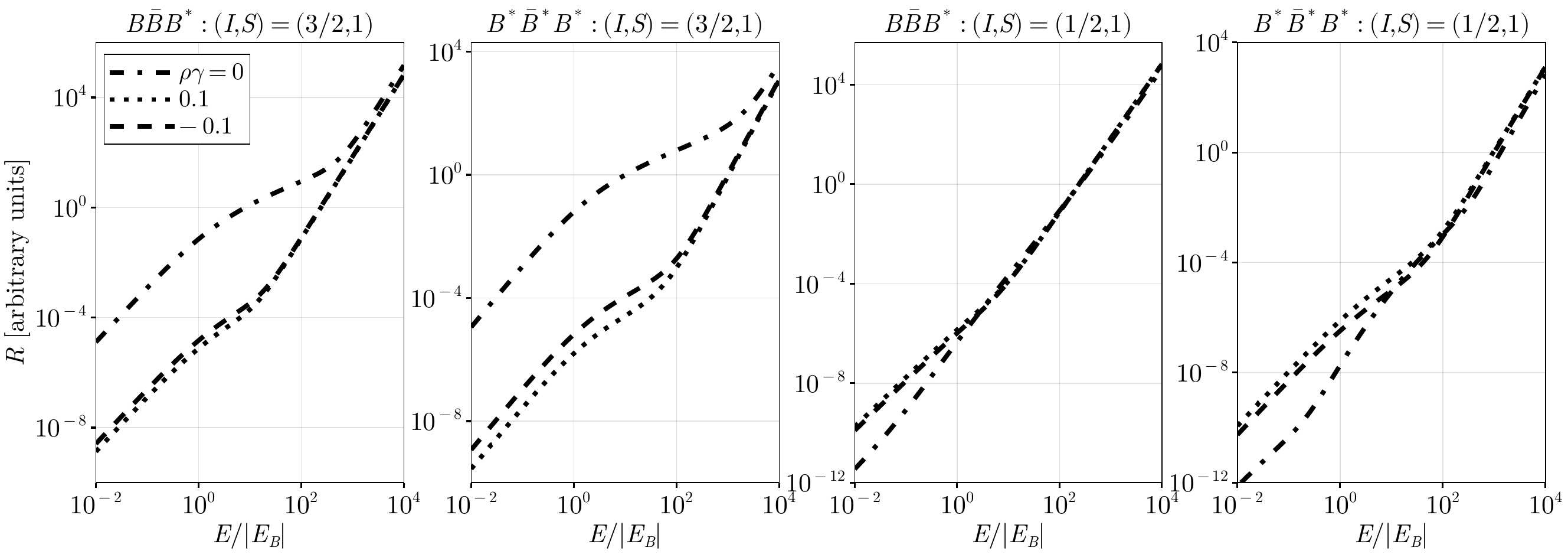}
    \end{center}
    \caption{\label{fig: rate3B_single_frc}{Finite-range corrections to the short-range production rates of the $B\bar{B}B^*$ and $B^*\bar{B}^*B^*$ three-body channels. The dash-dotted, dotted, and dashed lines represent the results for the zero-range case and for $\rho\gamma = 0.1$ and $\rho\gamma = -0.1$, respectively. 
    }}
\end{figure*}
\begin{figure*}[htb]
    \begin{center}
        \includegraphics[width=1\textwidth]{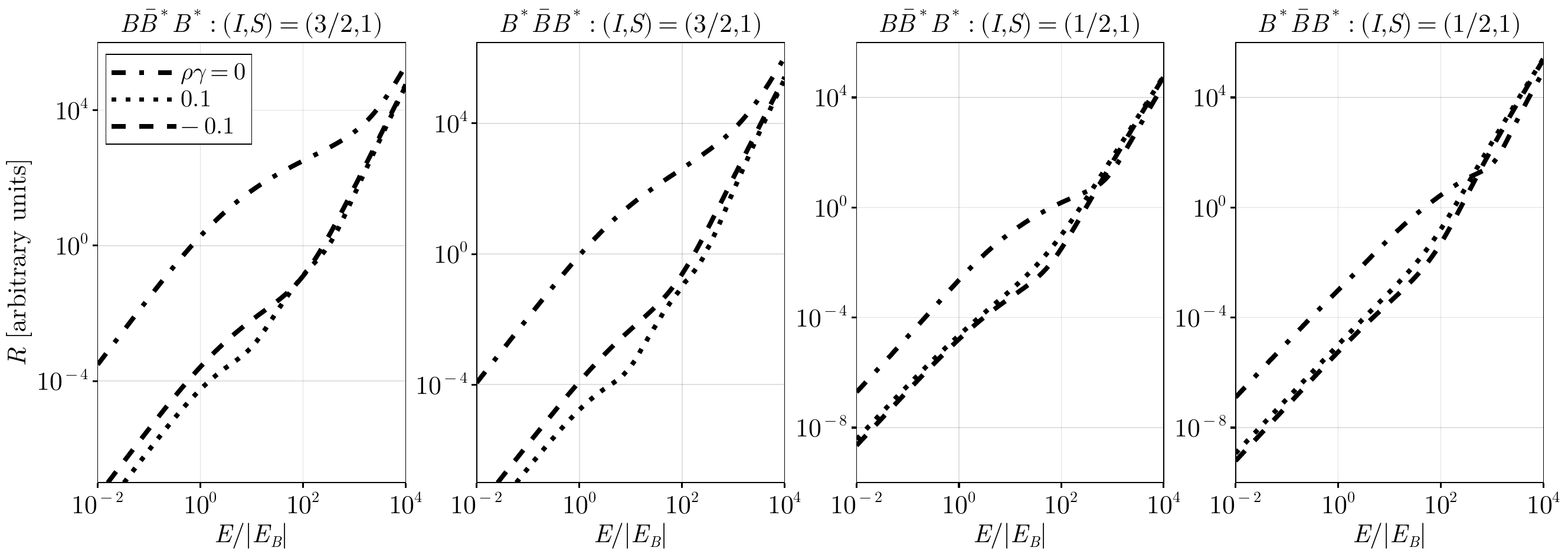}
    \end{center}
    \caption{\label{fig: rate3B_coupled_frc}{Finite-range corrections to the short-range production rates of the $B\bar{B}^*B^*$ and $B^*\bar{B}B^*$ three-body channels. The notations follow those in Figure~\ref{fig: rate3B_single_frc}. 
    }}
\end{figure*}

\section{Conclusion and outlook\label{sec:end}}
We extend our previous investigation of three-$B$-meson systems to explore the emergence of unparticles in their short-range creations. The absence of the Efimov effect in such systems, as established in an earlier study, indicates that the relevant observables can be reliably predicted within a short-range NREFT framework.
In this work, we calculate the short-range production rates for both the two-body dimer-particle channels $T_{b\bar{b}_1}B$ and the three-$B$-meson channels with quantum numbers $(I,S)=(3/2(1/2),1)$ at leading order in the short-range NREFT.

The observed power-law asymptotic behavior of these rates provides clear evidence for the emergence of unparticle dynamics in the three-$B$-meson sector, including the $B\bar{B}B^*$, $B\bar{B}^*B^*$, $B^*\bar{B}B^*$, and $B^*\bar{B}^*B^*$ unparticles governed by the underlying nonrelativistic conformal field theory.
Experimental confirmation of these characteristic power-law increases would offer strong evidence for large scattering lengths in the two-$B$-meson interactions, thereby providing a natural explanation for the dominant $B^*\bar{B}^*$ and $(B\bar{B}^*+\bar{B}B^*)/\sqrt{2}$ components in the internal structures of the $Z^\prime$ and $Z$ state, respectively.
Furthermore, finite-range corrections to the three-$B$-meson production rates, evaluated for $\rho\gamma=\pm0.1$, reveal a pronounced sensitivity to the effective-range parameters in the low-energy region, while preserving the same asymptotic power-law behavior at high energies.

Prompt production has previously been used to study the properties of the $X(3872)$ at the hadron colliders Tevatron and LHC \cite{Bauer:2004bc,Artoisenet:2009wk,ATLAS:2016kwu}, and can also provide insights in the bottom meson sector \cite{Guo:2013ufa}.
Future experimental determinations of the three-$B$-meson and $ZB$ short-range production rates considered here would thus provide valuable insights into the underlying two-body dynamics and the nature of the $Z$ and $Z^\prime$ states.
Moreover, these measurements provide a way to test the approximate conformal symmetry predicted for such systems \cite{Hammer:2021zxb,Braaten:2023acw,Braaten:2024tbm} at low energies experimentally.

\acknowledgments
H.-W.H. was supported by Deutsche Forschungsgemeinschaft (DFG, German Research Foundation) under Project ID 279384907 – SFB 1245 and by the German Federal Ministry of Research, Technology and Space (BMFTR) (Grant No. 05P24RDB).
U.-G.M. was supported in part by the Chinese Academy of Sciences (CAS) 
President’s International Fellowship Initiative (PIFI) (Grant
No. 2025PD0022), by the MKW NRW under the funding code NW21-024-A,  by the Deutsche Forschungsgemeinschaft (DFG,German Research Foundation) as part of the CRC 1639 NuMeriQS – project no. 511713970, and by the European
Research Council (ERC) under the European Union's Horizon 2020 research
and innovation programme (grant agreement No. 101018170).

\bibliographystyle{JHEP}
\bibliography{refs}

\end{document}